# Single-camera Two-Wavelength Imaging Pyrometry for Melt Pool Temperature Measurement and Monitoring in Laser Powder Bed Fusion based Additive Manufacturing

Chaitanya Krishna Prasad Vallabh and Xiayun Zhao*

ZIP-AM Lab, Department of Mechanical Engineering and Materials Science,
University of Pittsburgh, Pittsburgh, Pennsylvania, 15213

**Abstract**

Melt pool (MP) temperature is one of the determining factors and key signatures for the properties of printed components during metal additive manufacturing (AM). The state-of-the art measurement systems are hindered by both the equipment cost and the large-scale data acquisition and processing demands. In this work, we introduce a novel coaxial high-speed single-camera two-wavelength imaging pyrometer (STWIP) system as opposed to the typical utilization of multiple cameras for measuring MP temperature profiles through a laser powder bed fusion (LPBF) process. Developed on a commercial LPBF machine (EOS M290), the STWIP system is demonstrated to be able to quantitatively monitor MP temperature and variation for 50 layers at high framerates (> 30,000 fps) during a print of five standard fatigue specimens. High performance computing is employed to analyze the acquired big data of MP images for determining each MP's average temperature and 2D temperature profile. The MP temperature evolution in the gage section of a fatigue specimen is also examined at a temporal resolution of 1ms by evaluating the derived MP temperatures of the printed samples' first, middle and last layers. This paper is first of its kind on monitoring MP temperature distribution and evolution at such a large, detailed scale for longer durations in practical applications. Future work includes MP registration and machine learning of "MP-Part Property" relations.

*Author to whom the correspondence should be addressed - email: xiayun.zhao@pitt.edu; phone: (412)-648-4320

# 1 Introduction

In laser powder bed fusion (LPBF) based additive manufacturing (AM) processes, melt pool (MP) dynamics will critically determine the properties of printed parts and can serve as one of the key indicators for part quality [1-3]. MP morphology and temperature profile can help predict the process behavior and part condition. The major challenges in in-situ sensing and measurement for metal AM is the fast and complex process dynamics, which demand multiple sensors and high sampling frequencies for data acquisition and processing. Ultrahigh-speed synchrotron x-ray imaging has been adapted to study detailed MP dynamics and unveil important phenomena such as keyhole formation during laser melting of metals as practiced in LPBF [4-7]; however, such X-ray based instrument is extremely sophisticated and expensive limiting the accessibility to most AM researchers and industrial users for regular process monitoring. Other existing in-situ monitoring methods for metal AM processes typically adopt camera imaging methods to infer geometric flaws and or observe process anomalies (e.g., spattering and powder spreading issues) [8-11]. These methods have limited accuracy and interpretability due to the lack of information corresponding to the underlying thermal physics, which cannot be observed by simply using an in-situ camera but greatly accounts for the MP morphology and powder motion due to the temperature-dependent recoil pressure [12, 13], phase transformation and microstructure formation [14-16], as well as thermal residual stress-induced geometric flaws [17, 18]. In particular, there is a scarcity of affordable and capable measurement tools for tracking the in-process ultrafast MP temperature profile and evolution throughout an entire build.

In-situ temperature measurement is the key to achieve a comprehensive and insightful in-situ process monitoring for metal AM and understand the underlying physics causing the temperature related process phenomena and part properties. However, currently, there are limited technologies for monitoring and measuring the full-field temperature profile throughout the process at a sufficiently small-time scale (order of μs) to capture the ultrafast melting and cooling process during laser metal AM processes. Typically, a metal AM process involves MP, plasma, metal vapor, and spatter. Measuring the transient temperature of the MP is important for understanding the process dynamics such as the keyhole mode which accounts for keyhole pores generation. Also, it can provide information on heating and cooling rates, which significantly affect the microstructure. However, the ultrafast laser melting and solidification process occurs on the order of hundreds of microseconds (e.g., 200-600 µs) [19], necessitating a high-speed measurement. Besides, plasma, which is generated by laser irradiating the vaporization (caused due to the process), fluctuates with laser energy density and significantly affects the radiation by reflection or refraction [20]. This adds complexity to the temperature measurement.

Conventional methods such as photodiode, thermocouples, infrared (IR) cameras, and off-the-shelf pyrometers cannot measure absolute temperature locally, and accurately. For instance, the photodiode method can only provide single-point light emission signal which cannot be used to infer real temperature value or profile. Commercial IR thermography or pyrometers exhibit notable errors and uncertainties as the theoretical assumptions will be weakened in metal AM environment (e.g., spatter, vapor). Recent study using a commercial multi-wavelength pyrometer to monitor electron-beam metal AM shows good accuracy, but measures preheat temperature of powder bed (rather than MP) locally (at a single point only) at slow speeds (0.125 – 23 Hz) [21]. Emerging multi-wavelength imaging pyrometry has been developed to measure temperature in other fast processes such as combustion [22] and fire flames [23]. For metal AM processes, pioneering efforts in estimating the MP temperature profile were developed by researchers at KU Leuven [24-26], following which researchers developed systems using the two -wavelength pyrometry principles [27, 28]. In [24-26], the developed systems were co-axial monitoring systems, typically consisting of a high-speed CMOS camera and a photodiode, to capture the radiation emitted by the MP. The radiation captured by the two sensors was integrated to represent the MP intensity. The developed system was also employed to estimate the MP shape, process defects (such as balling, pore formation), and the MP temperature. However, the system's data acquisition speed was limited to 10 kHz, based on the scan speed and the scan width. This acquisition speed was achieved by reducing the field of view to a limited region of interest (< 60 × 60 pixels). The pyrometry systems employed in [27, 28] typically consisted of two photodiodes which register the surface thermal radiation and estimate the MP temperature in LPBF AM. Doubenskia et al [28] improved the system reported in [27] by using two InGaAs photodiodes, which helped increase the field of view of the system from 560µm to 1120µm. These systems provided good temporal resolution of the MP temperature readings without prior information of the emissivity values but were limited to a fixed region of interest (i.e., a single point measurement – cannot provide spatial profile). Recently, Hooper [29] developed a two-color high-speed (100,000 fps) based on the two-wavelength imaging pyrometry (TWIP) method, similar to the methods developed by Van Gestel [30] and Dagel et al [31] with improved hardware and software capabilities of the system, facilitating a higher frame-rate data acquisition and a practical application for rectangular test coupon for evaluating the MP heating and cooling rates. However, the recording times were limited to around 3 seconds for this test case. Other works to evaluate the MP temperature include coaxial high-speed camera-based temperature intensity measurements and infrared camera integrated with single high-speed camera (emissivity dependent) -based methods [11, 32]. Specifically, the MP temperature levels were estimated using a single wavelength MP emission radiance obtained from the co-axial camera [11]. In [32] the temperature measurements and cooling rates were estimated for single track measurement on IN-625 bare plates, using both high-speed camera (10,000 fps) and IR camera sensors. This work estimates the temperature based on the estimated emissivity values. All these methods provide valuable insights to the metal AM process and the MP temperatures. However, these methods are limited in measurement speed, monitoring duration, online data acquisition and storage capacity, data transfer rate, computational power for image processing and analysis, and equipment cost, hindering a wide adoption of MP temperature monitoring methods for research and process control of metal AM.

To fill the aforementioned gaps of in-situ MP temperature measurement and monitoring, in this work, we present a single-camera two-wavelength imaging pyrometry (STWIP)system and the associated methodological framework, which features a single high-speed camera as opposed to the usage of multiple cameras (conventional approach) to significantly reduce not only the equipment cost but also the amount of real-time image data.

Reduced data size greatly mitigates the issue of otherwise extremely demanding image storage and transfer capability, allows for extended monitoring time, and eases the workload of data processing. Moreover, the STWIP system's in-situ instrumentation is considerably compact, clearing one of the key roadblocks to install an in-situ MP monitoring system on various metal AM machines, especially on existing commercially available AM systems which are usually closed or strictly confined. As such, the STWIP method will enhance the AM community's accessibility to the much-needed online MP monitoring data, especially for a practical printing job that usually lasts much longer than the typical recording duration of a high-speed camera, providing valuable insights on process dynamics for research and development of metal AM, especially for the LPBF technology.

Although, commercial single sensor based two-wavelength systems exist (e.g., ThermaViz® from Stratonics Inc.), no literature reports have been found on using these systems to acquire and analyze high-speed and large-scale data for continuous MP temperature measurement during LPBF AM. Besides, these commercial systems are typically installed off axis with a very limited field of view on the build plate.

In this reported system, the camera and the optics system are coaxially aligned to the source laser and the reflected MP. The MP of a print is continuously monitored at framerates 30,000 fps and higher. The camera sensor provides an optical resolution of 20 μm/pixel. The intensities of the MPs acquired at the two working wavelengths are then analyzed for obtaining the MP temperature profiles. Experimental data and results show a great potential of the developed STWIP method for estimating MP temperature profiles accurately, along with the system's capability of continuously monitoring MP morphology and motion. The derived MP morphology, intensity and temperature can be further used to develop correlation models corresponding to the process-structure-property in LPBF and other metal AM processes.

The remainder of the paper is organized as follows, Section 2 details the theory of two-wavelength pyrometry, followed by the Materials and Methods (Section 3) including the STWIP system design and methodological framework, i.e., the optical system setup and characterization, image data processing and analysis method for temperature calculation, experimental validation, along with a quantitative analysis of measurement uncertainty. Section 4 and Section 5 discuss the experimental setup and related results. Section 6 concludes this work with recommendations for future work.

## 2 Theory: Two-Wavelength Pyrometry

The spectral radiance ($L_B$) of a blackbody (emissivity, $\varepsilon = 1$) is given by Planck's law:

$$L_B(\lambda, T) = \frac{2hc^2}{\lambda^5 \left(\exp(hc/k_B \lambda T) - 1\right)} \tag{1}$$

where $h$ is the Planck's constant, $c$ is the speed of light, $\lambda$ is the wavelength, $T$ is the absolute temperature of the body and $k_B$ is Boltzmann's constant. In the reported application, the printed parts or powders are not perfect blackbodies therefore the above equations have to be modified accordingly i.e., for a material with $\varepsilon \neq 1$, and by applying Wien's approximation, Equation 1 can be rewritten as:

$$L(\lambda, \mathrm{T}) = \frac{(2hc^2)\varepsilon A}{\lambda^5 (\exp(hc/k_B \lambda \mathrm{T}))} \qquad (2)$$

where $A$ is the transmission efficiency of the optical path. In the above equation the spectral radiance, $L$, is synonymous to the intensity measurement $I(\lambda, \mathrm{T})$. The emissivity is dependent on the material and can vary with parameters such as temperature, wavelength, and surface finish of an object. Therefore, the exact emissivity is often hard to estimate even with proper calibration methods [33]. To overcome the issues with emissivity measurements, the intensity measurements are often acquired at two different wavelengths.

Möllmann et al. [34] present a detailed discussion about the temperature measurement principle using two-wavelength ratio based thermal imaging, including both a theoretical analysis based on thermal radiation laws and practical implementations that involve the camera-related parameters (e.g., spectral sensitivity and filter behavior). According to their report, the selection of the two spectral bands is one of the most important parameters that influence the accuracy of the temperature measurement using an "emissivity-free" radiation ratio thermometry. One can use the wavelength and object temperature dependent value of $\frac{\partial^2 L(\lambda, T)}{\partial \lambda \partial T}$ to obtain the optimum wavelength region for the two-wavelength measurement, given the range of target object temperature. In our application case of LPBF monitoring, the MP temperature is typically ~2000 K and could go up to approximately 4000 K depending on the material and process conditions.

For satisfying Wien's approximation (Equation 2), i.e., the assumption of shorter wavelengths $\left(\lambda \ll \frac{hc}{k_B T_{MP}}\right)$, our chosen wavelengths must be significantly lower than 3599 nm and 7199 nm. These values were estimated from the above stated relation by considering the $T_{MP}$ values of 4000 K and 2000 K, respectively. Further from the analysis of the second derivative of spectral radiance, i.e., $\frac{\partial^2 L(\lambda, T)}{\partial \lambda \partial T}$ it is found that the optimum range of $\lambda T_{obj} \approx 1500\ \mu m\ K$ [35]. This relation can be used as a guideline for selecting suitable working wavelengths for specific applications. Therefore, as the typical MP temperature lies between 2000 and 3000 K, the spectral band is estimated to be from 750 nm (for 2000 K) to 375 nm (for 4000 K). Therefore, the optimum wavelength, based on the $\lambda T_{obj}$ value, lie between 750 nm and 500nm. Moreover, the wavelengths ($\lambda_1$ and $\lambda_2$) should be chosen to be close to each other (within 150 nm), which leads to the assumption of emissivity values ($\varepsilon_1$ and $\varepsilon_2$) being almost equal. Based on this theoretical analysis and the practical factors, such as the machine configuration, printing material's spectral intensity, we finalize the system's working wavelengths, as detailed in Section 3.1.

To conclude, the expression for calculating the temperature at the two measured intensities with the employed two-wavelength methodology is given by:

$$T = \frac{\frac{hc}{k_B}(\frac{1}{\lambda_2} - \frac{1}{\lambda_1})}{\ln\left(\frac{I_1}{I_2}\right) - \ln\left(\frac{\varepsilon_1}{\varepsilon_2}\right) - \ln\left(\frac{A_1}{A_2}\right) - 5\ln\left(\frac{\lambda_2}{\lambda_1}\right)} \tag{3}$$

Based on the assumption of $\varepsilon_1 \approx \varepsilon_2$, the term $\ln\left(\frac{\varepsilon_1}{\varepsilon_2}\right)$ is neglected in this work and the temperature is calculated independent of the emissivity values. However, this assumption will be further validated in our future works by independently validating the temperature measurements for different powder cases and an appropriate $\varepsilon_1:\varepsilon_2$ will be implemented.

## 3 Single-camera Two-wavelength Imaging Pyrometry (STWIP)

### 3.1 System Design and Setup of coaxial high-speed single-camera two-wavelength imaging pyrometry (STWIP) for MP temperature measurement and monitoring in LPBF

The in-situ coaxial high-speed STWIP system is installed on a commercial LPBF printer (EOS M290 DMLS). The schematics and the physical setup of our STWIP system is shown in Figure 1. As described in the previous section, the system is built on the principles of two-wavelength pyrometry. The key components of the STWIP system include, wavelength specific optical band-pass filters, achromatic doublets, beam directing optics and a high-speed camera (Nova S12, Photron USA, San Diego, USA). All the optical components were purchased from Thorlabs unless specified.

Our developed system is assembled to the EOS system at the beam splitter junction of the operating laser and the reflected MP radiance (Fig. 1). The incoming MP radiance is attenuated by the printer's beam splitter to a safe working power to ensure that the camera sensor and the other optical components are not damaged. The specifications of the beam splitter are proprietary. The subsequent light path is then divided into two separate arms for imaging the MPs at two separate wavelengths. The incoming MP radiance from the system is passed through a 50:50 beam splitter as shown in Figure1. Following which the MP radiance is filtered using the specific working wavelengths ($\lambda_1$ and $\lambda_2$) for the two-wavelength system (Equation 3).

The working wavelengths of the system are chosen as per the theoretical analysis detailed in Section 2, which recommends an optimum spectral band between 500 nm and 750 nm, as well as the actual system's spectral response. Reflected spectrum for printing standard AM powders were acquired (Inconel 718 and Stainless Steel 316L) using a spectrometer (BTC112E, B&W Tek, Newark, DE), immediately after the EOS beam splitter (before our optical system). From these spectra, we observed two peaks around 550 and 620 nm respectively, a representative spectrum for IN718 powder printing is shown in Figure S1 in the Supplemental Information (SI). Note that the built-in beam splitter on the EOS machine attenuates the spectrum above 750 nm by design (See Fig. S1), significantly limiting the lowest detectable temperature to ~1300 K as observed in our experiments for typical print parameters. Given a wider spectrum (beyond 750nm) that is likely available from other machines such as an open-architecture LPBF, the working wavelengths can be chosen to be near infrared or longer to measure lower temperatures (e.g., below 1000 K) thus extend the temperature measurement range. Based on the cumulative reflective spectra for these materials and the spectral cut-off of the EOS printer, the working wavelengths of our system are chosen to be 550nm ($\lambda_1$) and 620nm ($\lambda_2$), which satisfy both the theoretical (Wien's approximation and the optimum wavelength-temperature relation, *$\lambda T_{obj} \approx 1500$ μm. K*) and practical conditions.

Each light path is equipped with iris apertures for independent aligning and testing. Further, each wavelength arm is also equipped with achromatic doublets to help focus and image the two MPs on the same plane. The two wavelength arms are merged at the polarizing beam splitter (to remove any polarization effects and attain less distorted wave fronts) and are then spatially separated by precisely deflecting the kinematic mirrors. Finally, the two MP images are imaged at the camera sensor using a focus lens (SP 60mm f/2, Tamron USA, Commack, NY). It is worth noting that the developed system is designed to be modular, making it readily applicable to different measurement scenarios. For instance, the systems spectral filters and the other optics can easily be switched with an additional utility of adding an additional optical component for additional measurements (such as a photodiode or an infrared camera or even another wavelength branch system to form a multispectral imaging system).

Using the developed optics setup, the MP data at two wavelengths are acquired during the print process. Representative MP images are shown in Figure 2. These MP images correspond to the Fatigue Test Samples print. The data was acquired at 30,000 fps with a resolution of 128 × 48 pixels at an optical resolution of 20μm/pixel. For estimating the MP temperature accurately, optical path transmission efficiency terms $A_1$ and $A_2$ shown in Equation 3 must be determined. The $A_1/A_2$ ratio analysis is detailed in the next section.

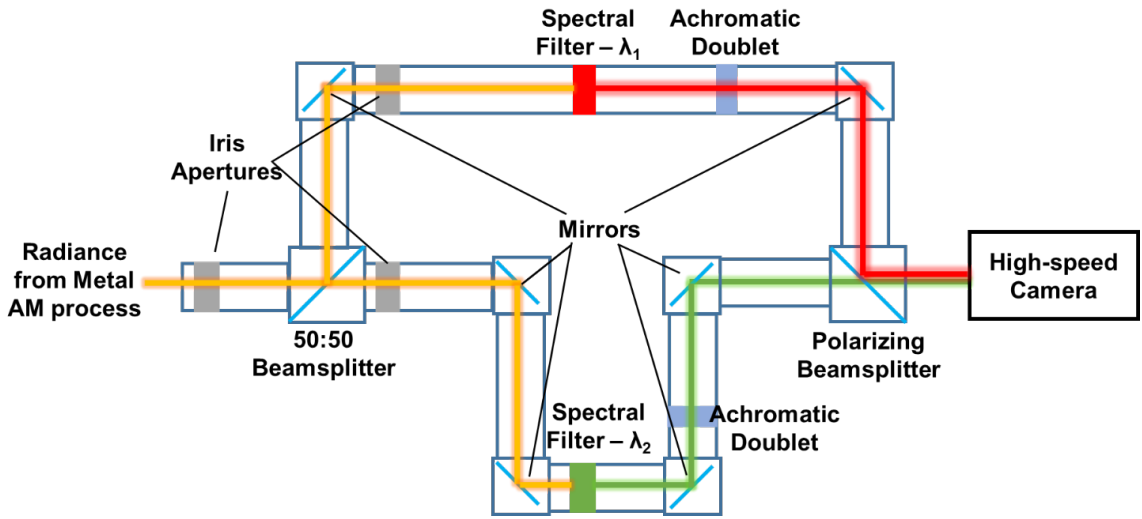

(a)

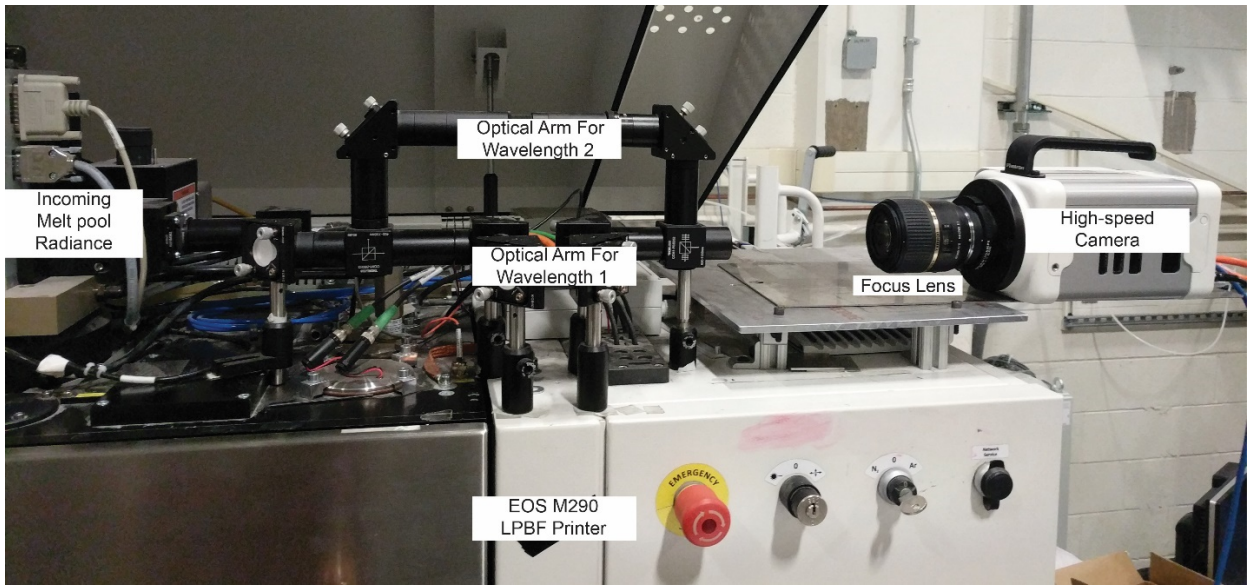

(b)

Figure 1: The developed system of single-camera two-wavelength imaging pyrometry (STWIP) for melt pool temperature measurement and monitoring during LPBF based metal AM: (a) schematics; and (b) physical system deployed on the EOS M290 LPBF printer.

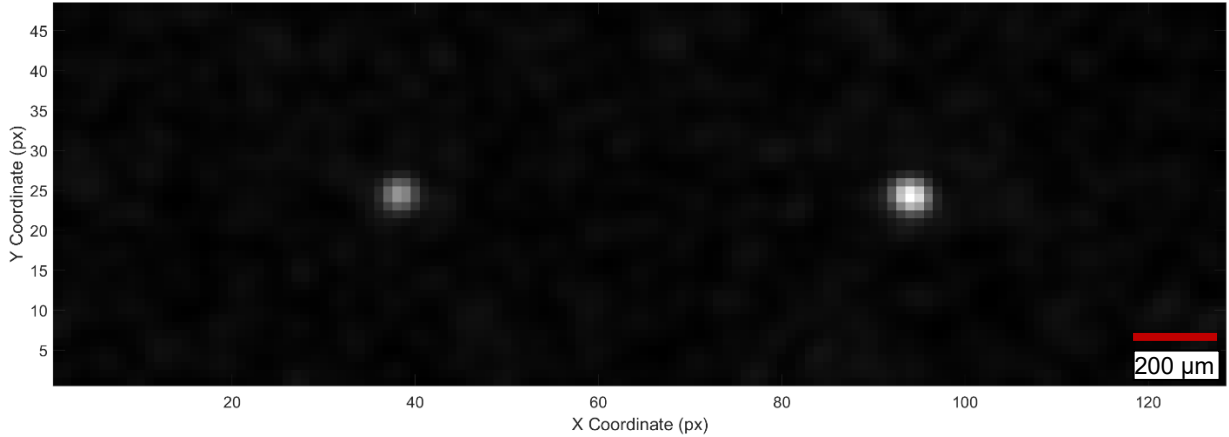

Figure 2: Representative in-situ STWIP acquired raw data of two-wavelength melt pool images during a fatigue specimen printing process (resolution: 128×48 pixels; acquisition rate: 30,000 fps). The sensor has an optical resolution of 20μm/pixel. The melt pool on the right corresponds to 550 nm and the melt pool on the left corresponds to 620 nm.

### 3.2 Optical system calibration to characterize the two-wavelength optical transmission ratio $A_1/A_2$ ($A_{12}$)

For evaluating the $A_1/A_2$ ratio (or $A_{12}$) in Equation 3, the transmission efficiency of the optical path corresponding to the two working wavelengths needs to be characterized. For this, a broadband light source is placed in the build chamber and the light intensity at the source and at the end of the optical path are acquired using a spectrometer (BTC112E. B&W Tek, Newark, DE) (Figure S2). The ratio of the intensity values at the working wavelengths, i.e., 550 nm ($\lambda 1$) and 620 nm ($\lambda 2$) are calculated which are then substituted in Equation 3 to evaluate the temperature value for a given intensity ratio value. Representative plots of the source spectrum and the spectrum at the end of the optical path are shown in Figure S3. The spectra are acquired for multiple test cases to account for measurement errors. Besides, an extensive study with repeated optical calibration experiments is performed using a sturdy spectrometer setup, (1) to mitigate the consequence of potentially inconsistent measurements of light intensity input to the EOS machine's F- θ lens; (2) to evaluate and verify the accuracy and precision in the measurement of optical transmission efficiencies, $A_1$ and $A_2$, for the two-wavelength arms in the system; (3) to understand the effect of build location on the two-wavelength transmission ratio $A_{12}$, and (4) ultimately, to enhance the accuracy and robustness in temperature measurements via STWIP.

Ten individual measurements of $A_{12}$ are acquired at five representative locations (i.e., a total of 50 individual measurements) on the build plate, for evaluating the actual $A_{12}$ of our developed STWIP system (refer to the SI Section II for details). It is found that the deviation in the ratio is within ±0.0163 for all the tested locations (refer Fig. S2b). Therefore, we conclude that the effect of build location on the $A_{12}$ is minimal. The cumulative $A_{12}$ ratio of the system is determined to be 1.601 ± 0.0163 (see Table S1 and Fig. S5). By substituting the derived $A_{12}$ into the Planck's law (Equation 3), we determine the temperature-intensity ratio ($I_1/I_2$ or $I_{12}$) relation as shown in Figure 3. The MP temperatures in the subsequent experiments will be evaluated based on this relation. Specifically, the MP image intensity ratio will be obtained from the MP images for each optical wavelength (550 and 620 nm), the obtained ratio will be substituted into the Equation 3 along with the characterized $A_{12}$ for determining the MP temperature. For accurately determining the Intensity ratio of these MP images, both the images must be accurately aligned (i.e., same spatial coordinates). The alignment of these images is achieved by using a feature recognition algorithm as detailed in the next section.

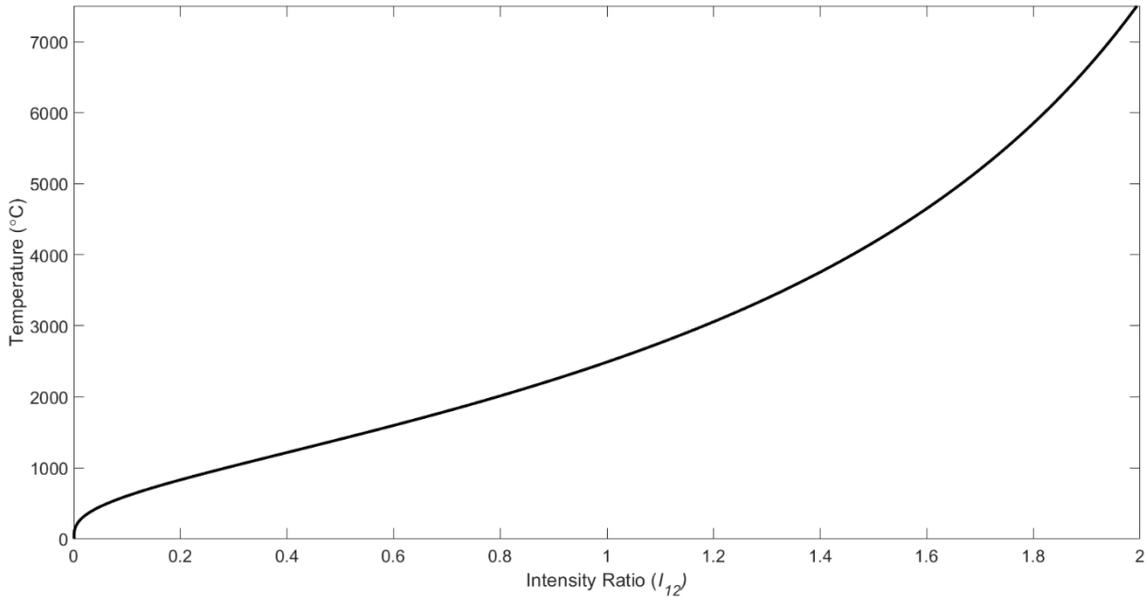

Figure 3: Temperature evaluated for different values of the two-wavelength intensity ratio $I_{12}$, using the calibrated $A_{12}$ of 1.601±0.0163 based on the simplified Planck's law (Equation. 2)

### 3.3 Two-wavelength melt pool images processing to extract the two-wavelength intensity ratio ($I_{12}$) and calculate MP temperatures

Due to the possible minor misalignments in the optical system setup, and the wavelength-dependent optical wave propagation of the MPs and the probable aberrations in the optical path, the MP images acquired by the camera sensor corresponding to the two working wavelengths of the system may not be identical in terms of size and orientation. In order to determine the intensity ratio, thus each MP's temperature distribution from its corresponding pair of two-wavelength images, the images of one working wavelength must be appropriately scaled and transformed with respect to the images of the other working wavelength for ensuring an exact spatial coordinate matching (or mapping) between the two-wavelength images for extracting the accurate MP intensity ratio ($I_{12}$) information.

In conventional two-camera or higher number of camera based two/multi-wavelength imaging pyrometry systems, the image alignment is typically performed by both hardware alignment and wavelength specific offset and scaling corrections using image processing software. The focus of the MP image at each wavelength could be individual aligned by appropriately adjusting each camera's focus and position to achieve a good match of the MP's images at different wavelengths. However, there is a possibility of error induced due to the use of different imaging sensors and the relative location of the MP on the location of the sensor. Even a difference of one-two pixels could cause an intensity ratio mismatch resulting in inaccurate temperatures. Also, for open-architecture LPBF systems, researchers can align the images from two cameras using the feedback signal (the translations in x and y position of the imaged MP are related to the angular movements of the galvanometer scanning mirrors) from the system's galvo mirror motors [29].

In our STWIP system, as both the MP emitted wavelengths are imaged on the same camera sensor using the same optical path, an imaging processing based software approach is more viable in contrast to the abovementioned hardware aided approaches. As the STWIP method has an advantage of circumventing the potential issues caused by multiple camera sensors, it also presents more challenges in image processing of the two-wavelength images. For instance, improper image transformation/alignment (scaling and rotation) of the two wavelength images could result in inaccurate intensity ratios, thus resulting in false temperature measurements. Therefore, a robust alignment/transformation procedure is recommended for achieving accurate and precise MP temperature measurements. Our methods for the image transformation are detailed below.

**Image Processing:** To accomplish the above mentioned image alignment and transformation, we employ a feature recognition algorithm, known as the KAZE algorithm [35] for extracting the scaling and rotation information of the MPs and transforming them appropriately. KAZE algorithm is a non-linear scale-space based method, which is invariant to scale, rotation, and is more distinctive at varying scales as opposed to other existing feature recognition algorithms [35, 36]. The KAZE algorithm is implemented using MATLAB. Here, we typically designate one MP image as the reference MP (say MP1 at 620 nm), then the second MP (MP2 at 550 nm) is compared with MP1 for detecting the features based on the intensity, edge, and boundary profile. KAZE algorithm requires at least three matching inlier points to successfully match and transform the images. Typically, KAZE algorithm is employed for feature recognition of macroscale images (256 x 256 pixels or greater). In this work, we applied KAZE to MPs which are typically on the order of 15x15 pixels. Initial application of KAZE to these images was not successful, resulting in a very low throughput. For improving the accuracy of the recognition and transformation, we up-scale the MPs (typically 3x) and then process them through the algorithm for feature recognition and transformation. The transformed images are then downscaled for the intensity ratio analysis.

To check the accuracy of the image transformation, the similarity or the matching factor of the transformed images was checked using a structural similarity index function (*SSIM*) in MATLAB (see SI Section III). The average matching metric of these MPs was around 85%. And, due to the size of MP images, and the varying intensity profile during the AM process, we could only achieve a successful transformation of about 75% (i.e., 75 out of 100 MPs were successfully transformed) using KAZE algorithm. The feature recognition and transformation algorithms can be improved in the future to achieve higher degree of accuracy and speed in transforming the images and estimating the temperatures.

A representative MP transformation example is illustrated in Figure 4. Figure 4a and b show the individual segmented MP images at two wavelengths (620 nm and 550 nm). The raw MP images are noisy (see Figure 2) and contain both MPs in a single image. Therefore, these images need to be separated and filtered for further processing. The raw images are segmented using standard image processing methods (adaptive threshold, masking, and contouring) in MATLAB. Applying KAZE to noisy images can lead to inaccurate scaling and rotation. From Figures. 4a and b it can be noted that some of the background noise pixel still exist after segmentation, further thresholding this image would result in the loss of MP boundary. The black background indicates the lack of MP related pixels, thus can be considered as pseudo zero intensity pixels.

In order to obtain the accurate intensity ratio, the spatial coordinates of both the MPs should be same. Figure 4c shows the original 620 nm image and the transformed 550nm image, in the same spatial domain with their respective pixel intensities. In Figure 4c, MP image corresponding to 550nm is transformed to match the counterpart, i.e., the reference MP image at 620nm using KAZE algorithm. It can be seen that the center (core) pixels of both images perfectly match in the spatial coordinates. The MP intensity variation among these two wavelength images can also be clearly seen from these images, for this work, typically, the MP images at 550nm have higher intensity values compared to the MPs at 620nm.

Similarly, as shown in Fig. 4, for each MP, KAZE is applied to process the pair of two-wavelength MP images and obtain the scale and rotation for the specific pair and transform one image to match the other.

Further, these transformed images are analyzed for obtaining their pixel-wise intensity ratio ($I_{12}$). The pixel-wise intensity ratio is then substituted in Equation 3 for obtaining a 2D spatially resolved temperature profile of the melt pool with an optical resolution 20μm/pixel. Note that the intensity, scale and rotation between the two-wavelength images are subject to change throughout an LBPF based AM process due to the print process parameters, hatching patterns, and other stochastic disturbances (e.g., deflection of processing laser, vibration of the platform that holds the measurement optics). Therefore, one cannot simply assume and use constant scale and rotation factors. Instead, each pair of melt pool images has to be processed for its specific scale and rotation and then transformed accordingly to estimate the intensity ratio and thus the temperature profile.

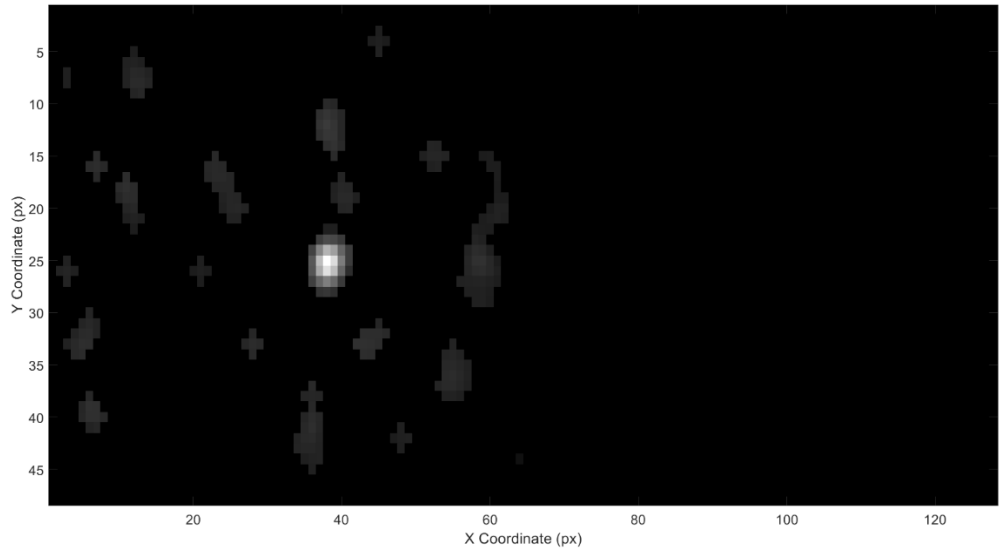

(a)

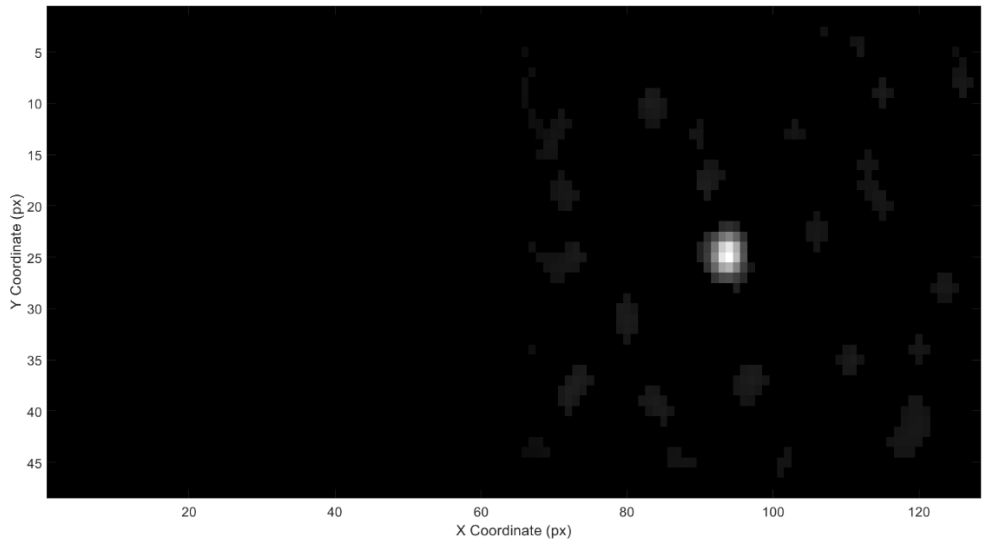

(b)

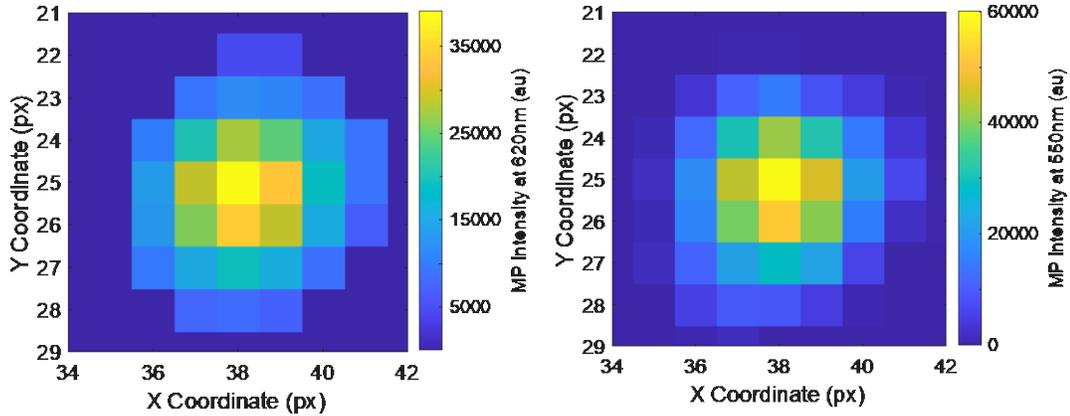

(c)

Figure 4 (a) and (b) show the segmented, individual melt pool images corresponding to 620 nm and 550 nm, respectively. Some background pixel-noise can also be observed in the segmented images. (c) shows the original 620nm melt pool image on left and the transformed 550nm on right. The X- and Y- coordinates for the two-wavelength images are the same after transformation, facilitating the pixel-wise intensity ratio measurements. The blue background in these images indicate the absence of melt pool pixels and represent pseudo zero intensity pixels [please refer to web version of the article for color images].

To validate the functioning of the employed feature recognition (KAZE) algorithm, a verification experiment was conducted using a light source with a known shape. This light source was placed in the EOS build chamber and was imaged using our STWIP system. The images and the analysis results are shown in the SI Section III (Figure S6). From this experiment, we can confidently establish the functioning of our employed featured recognition for the images obtained using our developed STWIP system.

**MP Temperature Calculation and Morphology Monitoring**: For this analysis, we first define a set of MP temperature signatures to be evaluated in this work, which include (1) 2D MP temperature profile, and (2) MP average temperature, with a prospect of future utilization in research on "MP - part property" relations.

For evaluating the temperature profiles, image transformation is applied to the segmented MP image as introduced above, the intensity profile for each image is extracted using the pixel wise intensity information. The acquired MP images are 12-bit images, which are up sampled to 16-bit using the high-speed camera specific software (PFV 4, Photron USA, San Diego, USA) for better intensity scaling. For these images, 0 indicates the lowest possible pixel intensity and 65536 the highest possible pixel intensity. All the image processing is performed in MATLAB. From these extracted pixel wise intensity profiles, the intensity ratio of the two MP images is determined i.e., MP image at 550nm is divided by the MP image at 620nm for extracting the $I_{12}$ pixel-wise ratio. The pixel-wise $I_{12}$ is substituted in then Equation 3 for extracting the detailed 2D temperature profile of the MP. If the transformed image is not a good fit to the reference image, the intensity ratio profile will result in impractical numbers resulting in unrealistic temperatures, causing measurement errors that may be observed occasionally during the implementation of STWIP for MP temperature measurement as shown in the subsequent sections.

For evaluating the average intensity profile and the average MP temperature profile of a printed layer, MP images are processed using a custom written code. The code detects the highest intensity pixel in the image. Following which, the pixels surrounding the maximum intensity pixel in both X- and Y- directions are checked for the lowest possible intensity for identifying the MP boundary. The pixel values in this boundary are then averaged for estimating the average MP intensity. The average intensity profiles corresponding to $\lambda_1$ (550nm) and $\lambda_2$ (620nm) are then divided for obtaining the average intensity ratio ($I_{12}$), which is then substituted in Equation 3 for deriving the average MP temperature profile.

Also, it is worth noting that the MP images can be directly analyzed to obtain information of MP morphology including MP size (area and width) and shape. The capability of monitoring MP morphology using the STWIP is elaborated in our other published work [37].

### 3.4 Validation of the developed STWIP system and method for MP temperature measurement

For validating the temperature measured by the developed STWIP system, we performed temperature validation experiments using high-temperature Type C thermocouples (XMO-W5R26-U-062-36-H-SX-8, Omega, Norwalk, CT) (Figure S7). The thermocouples were inserted into a build plate and the laser was scanned over the thermocouples in two distinct patterns (Fig. S7b). The thermocouples were placed approximately 80μm below the build plate surface such that the temperature of the MP could be acquired by the thermocouples at a minimal risk of being damaged. Also, this depth of 80μm was chosen based on the typical MP depth (~100μm) for typical print parameters for single track prints [13]. Two test cases with two different scanning patterns (Figure S7b) were performed in an effort to increase the chance for the thermocouple to capture the MP temperature, because the commercial machine does not allow users to directly shine the laser on a fixed spot (target spot of thermocouple sensing tip in this work) long enough on par with the thermocouple's 100% response time (~1.65 seconds). Our initial tests (Case #1) used a rectangular pattern (size 10×20 mm, scan speed 150mm/s, laser power 25W), and the next set of tests (Case #2) used an array of star-like pattern (3 mm diameter) within an area of 10×20mm (see Fig. S7b) (scan speed 50 mm/s, laser power 50W) to scan over the thermocouples. These settings were chosen to expose the thermocouple for at least one time constant of the thermocouple (0.33 s). In our experiments, we were not able to generate a stationary laser scan at a target MP location for more than 0.5 seconds due to the constraints of the commercial LPBF system. Therefore, the thermocouple temperature measurements were estimated based on the step change percentage of the thermocouple corresponding to the time constant for a Type C thermocouple (SI Section IV). These prints in the two test cases were monitored by using the developed STWIP system and commercial thermocouples simulataneously. The temperature measurement performance of the STWIP system and methods is then validated with the thermocouple responses.

As a result, the relative measurement difference between thermocouple and STWIP was found to be 12.08 ± 2.77% in test Case #1 and 2.29 ± 1.50% in Case #2. The stark contrast of the two cases reveals that the small star-like patterns (Case #2) resulted in a much closer agreement between the STWIP and thermocouple measurements, mainly because it scans smaller patterns close to the length scale of the thermcouple sensing tip and thus allows the thermocouple to sense more of the actual MP in our compromised experiment setup due to the machine constraints. Moreover, Case #1 displays a small standard deviation (2.77%) despite a large average value of relative STWIP-Thermocouple difference (12.08%), indicating a promising repeatability. Case #2 shows an even smaller standard deviation (1.50%), further confirming that STWIP has a great repeatability. Provided a machine that allows a desired setup of directly measuring MP with a thermocouple for sufficient time (as practiced in [29]), the comparison of the STWIP against a thermocouple will be more straightforward and can validate STWIP with more confidence.

Moreover, the observed differences between the thermocouple and STWIP measurements might partly stem from inherent systematic errors in both the systems (measurement accuracy of the thermocouple used is ~±1% (for temperatures 426 – 2300°C)). It is also speculated that the STWIP-Thermocouple measurement difference might reflect an actual difference between the STWIP measured MP surface temperature and the thermocouple measured sub-surface temperature, which could be 10's°C to 100's°C lower, depending on the location of thermocouple sensing tip under the MP [13] (note the thermocouple is embedded beneath the build plate by 80 microns and may not sense the actual MP point). Nevertheless, we adopted this cost-effective thermocouple approach to rapidly evaluate the performance of STWIP and demonstrate its excellent accuracy and precision to a reasonable degree. A more direct yet expensive validation method using a standard temperature calibrating lamp or infrared camera will be used to further evaluate the measurement accuracy of STWIP in the future. More details of the validation experiments are provided in SI Section IV.

## 3.5 Uncertainty Analysis

To understand the temperature measurements uncertainty of the developed STWIP method, we performed a theoretical uncertainty analysis for the temperature measurements based on Equation 3. Assuming zero uncertainty in the values of wavelengths - $\lambda_1$ and $\lambda_2$ - and the light speed $c$, the uncertainty of temperature mainly resides in the two-wavelength optical transmission ratio $A_{12}$ and intensity ratio $I_{12}$. The temperature measurement uncertainty $U_T$ can be estimated by Equation 4, where $U$ represents the uncertainty in the measurement of a quantity indicated by the corresponding subscript, for example, $U_{A_{12}}$ and $U_{I_{12}}$ means the uncertainty in measurement of $A_{12}$ and $I_{12}$, respectively. $U_{A_{12}}$ is determined in Section 3.2. The uncertainty in the intensity ratio can be obtained by $U_{I_{12}} \approx \sqrt{U_{camera}^2 + U_{Transform}^2}$, where $U_{camera}$ means the uncertainty in intensity measured by the camera sensor, and $U_{Transform}$ refers to the uncertainty caused by the two-wavelength images transform and mapping algorithms (Section 3.3). For the uncertainty in this work, we do not consider $U_{Transform}$ due to the complex behavior of the MPs and the transformation errors induced through the feature recognition algorithm. Based on the specification of our camera, we estimate our current STWIP system's $U_{I_{12}}$ is approximately (0.0003). A detailed analysis with example calculations and plots is presented in the SI Section V.

$$U_T = \sqrt{\left(\frac{\partial T}{\partial A_{12}} \cdot U_{A_{12}}\right)^2 + \left(\frac{\partial T}{\partial I_{12}} \cdot U_{I_{12}}\right)^2} \qquad (4)$$

Based on our analysis, the temperature measurement uncertainty at different levels of $I_{12}$ given the characterized values of $A_{12}$ = 1.601 with an uncertainty of $U_{A_{12}}$ = 0.0163 is elaborated in SI Section V with the corresponding uncertainty plots (Figs. S9 and S10). In the specific LPBF system monitored by our STWIP, the intensity ratio $I_{12}$ usually range between 0.85 and 1.5. For instance, at a typical value of $I_{12} = 1.1$ *as* observed in our experiment (see representative $I_{12}$ histogram in Fig. S11), the temperature uncertainty due to the uncertainty of temperature in $A_{12}$ is 31.7ºC. Similarly, the temperature measurement uncertainty at different levels of $A_{12}$ given the typical value of $I_{12} = 1.1$ with an uncertainty of $U_{I_{12}} = 0.0003$ is 0.85ºC.

By substituting the two uncertainty results above into Equation 4, the overall temperature measurement uncertainty of the reported STWIP method for a typical $I_{12} = 1.1$ is 31.7ºC, (i.e., 2760 ± 31.7°C, see SI Section V for details). Figure S10 shows the total uncertainty, $U_T$ vs *Temperature,* for a constant $A_{12}$ of 1.601 and $I_{12}$ values of 0.5-2. It is found that the relative uncertainty in temperature evaluated from our system is less than 2.8%. This relative uncertainty (2.8%) is the maximum relative uncertainty of the system evaluated at $I_{12} = 2$, whereas an $I_{12}$ ratio of 2 is very uncommon in typical print case scenarios.

To conclude, the uncertainty in temperature measurement due to the change in $I_{12}$ (for $A_{12} = 1.601$, actual value of our STWIP system) is relatively lower compared to that due to the change in $A_{12}$ (when $I_{12} = 1.1$, which is typical in our experiment (See SI Figure S9). This comparison implies that STWIP temperature measurement performance significantly depends on the two-wavelength MP image acquisition (e.g., well-aligned optical setup, spectral sensitivity of camera) to obtain intensity values and intensity ratios with desired accuracy and precision. Additionally, the image transformation also plays a crucial role in accurately evaluating the intensity ratio. This analysis also shows that the $U_T$ will be reduced as $I_{12}$ decreases and $A_{12}$ increases, indicating a direction of improvement to design a favorable optics system with higher $A_{12}$.

## 4 Experimental Case Study: MP Temperature Monitoring during LPBF of Fatigue Specimens

This section demonstrates the implementation and application of the developed STWIP system and method for MP intensity and temperature monitoring during AM of fatigue test specimens, referred as "fatigue bars", showing the advantage of STWIP over conventional methods.

### 4.1 Experiment Design and Data Acquisition

Five fatigue bars were printed with a commercial AM powder, Ni-Cr super alloy, also referred to as IN718, (VDM® Alloy 718, Werdohl, Germany) using an EOS M290 DMLS printer. These specimens were designed based on the ASTM E466 standards and printed using default print parameters for IN718, i.e., Laser Power of 285W, Laser Scan Speed of 960 mm/s, Hatch Spacing of 110 μm, and Stripe Width of 10 mm. The print schematic is shown in Figure 5a (see Fig. S12 for the actual printed parts). The print comprised a total of 100 layers with a nominal print layer height of 40μm, out of which we acquired the in-situ MP data for 49 layers. Although our STWIP system would significantly reduce (by half) the number of images generated in unit time and thus data transfer time in contrast to the state of art, the data for some intermediate layers could not be acquired due to the camera storage and data transfer limitations. Note in this work we were able to acquire MP data at 30,000 fps continuously for ~74 seconds within each monitored layer and for 49 layers with occasional between-layer stops due to data transfer need, during the print of 5 standard fatigue bars, while current research using two-camera two-wavelength imaging pyrometry could monitor just over 3 seconds at 100,000 fps for a simple, rectangular test geometry [29]. To further increase the monitoring duration and continuity, we can improve our hardware and software capabilities in STWIP to increase the data acquisition capacity by four times.

Specifically, for this dataset, the data was acquired from Layer 14, the first 13 Layers are machining stock added for machining the printed part off the build plate using Electric Discharge Machining (EDM). The data was acquired in sets of three consecutive layers with a data transfer break (about the print time of two layers) between each two adjacent sets. That is, after the data for Layers 14, 15, and 16 were acquired, the data had to be transferred to an external solid-state drive, which typically elapses 60 – 70 seconds depending on the data size. To maintain uniformity, two layers for every three layers were skipped. Therefore, Layers 17 and 18 were skipped and the data acquisition resumed from Layer 19 to Layer 21 then skipping two layers. Due to data loss (incomplete recording, missing frames etc.) in some layers, the data corresponding to those layers (Layers 36, 45, 56 and 96) are not reported.

The MP image data was acquired using the high-speed camera (NOVA S12, Photron) at a frame rate of 30,000 fps and a pixel resolution of 128 × 48 pixels (Figure 2). Each Layer's data was acquired with the same camera settings. Each print layer elapsed about 72 seconds resulting in approximately 2.2 million frames per a print layer. The total size of the dataset was close to 1TB. The data was converted from video format to frame-by-frame image format, which were then processed in MATLAB for estimating the MP intensity and thereby temperature profile. Due to the large data set size, the data was analyzed on a supercomputing platform (Bridges-2, Pittsburgh

Supercomputing Center) using high performance computing clusters (HPCs). The intensity and the temperature profiles of these MP images at the two working wavelengths are calculated using the methods detailed in Section 3.

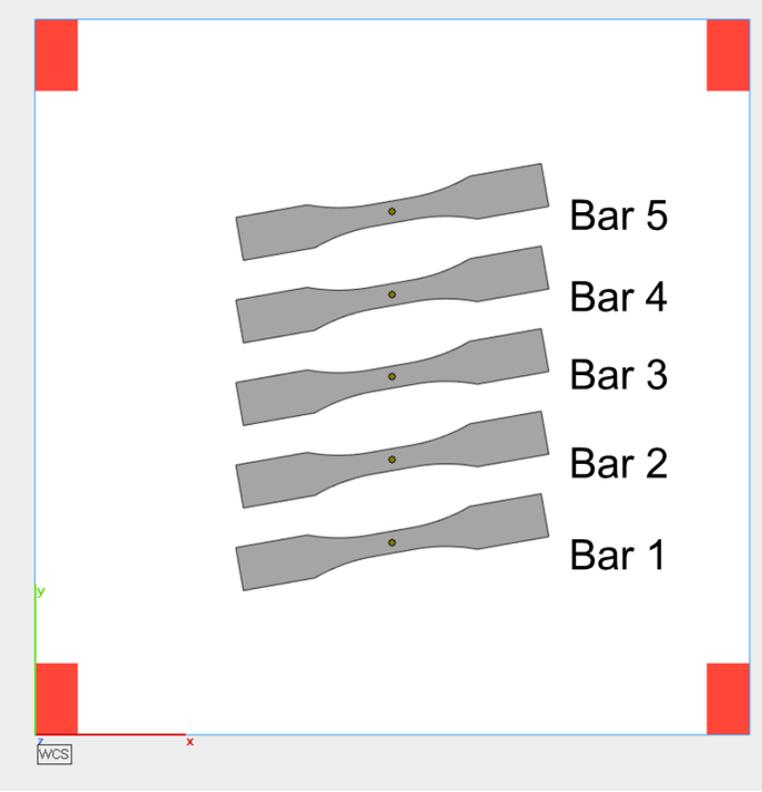

(a)

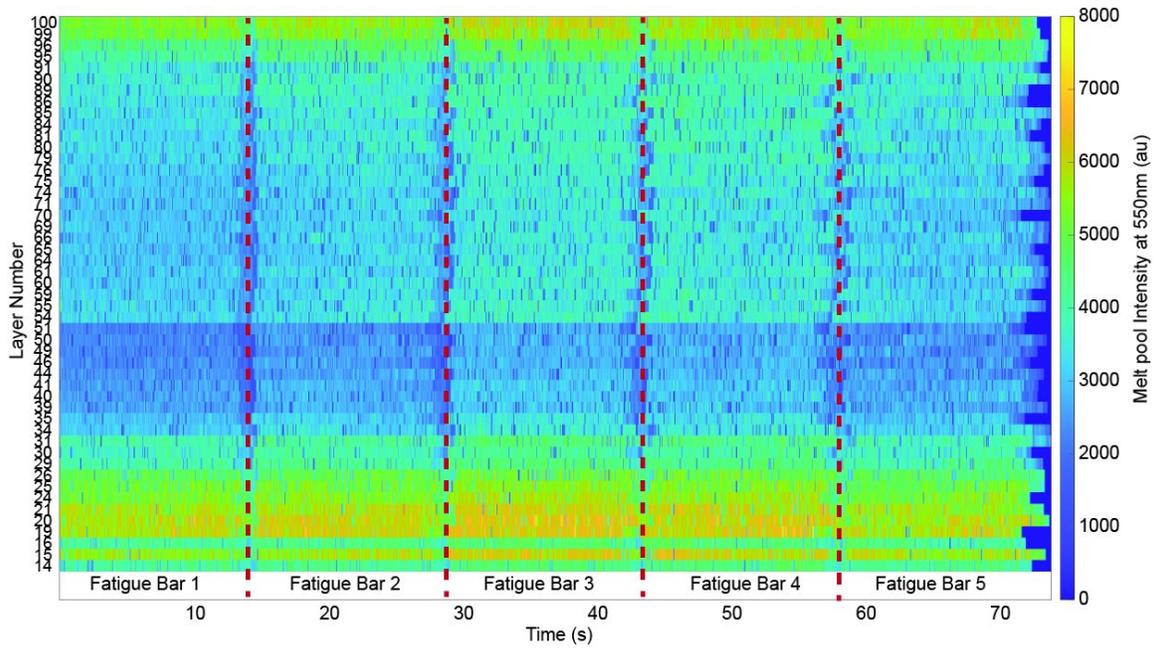

(b)

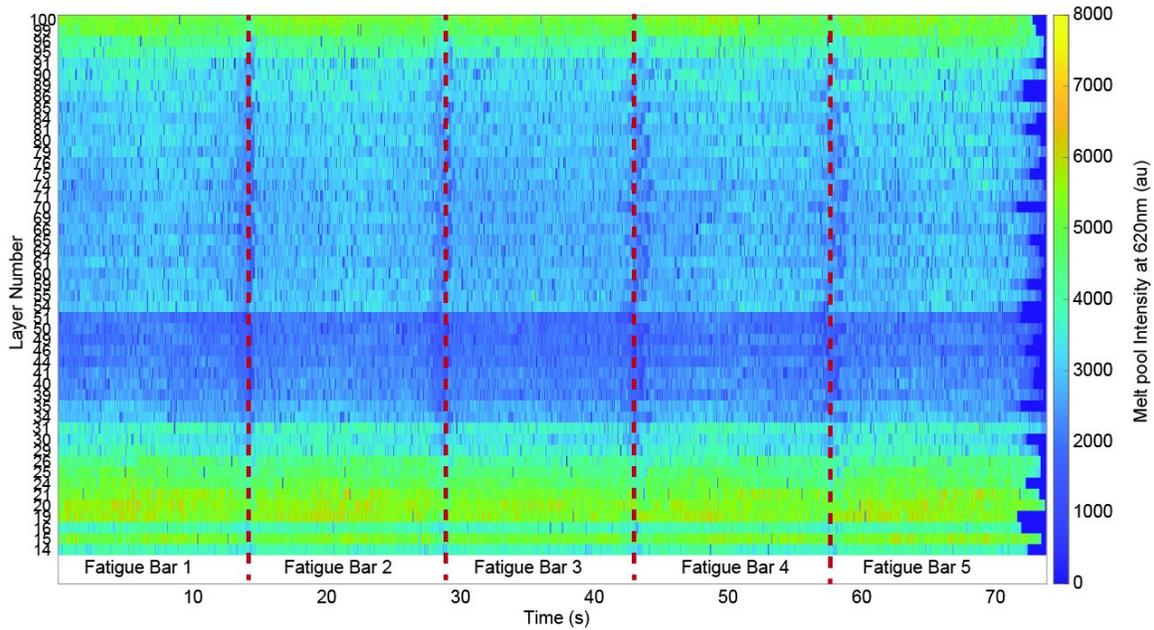

(c)

Figure 5 (a) Schematic of the Fatigue Test Samples print (b) and (c) MP intensity profiles of the five printed Fatigue Bars for 49 layers corresponding to (a) 550 nm and (b) 620 nm. The red dashed lines show the approximate the temporal boundaries of each printed Fatigue Bar. [please refer to web version of the article for color images]

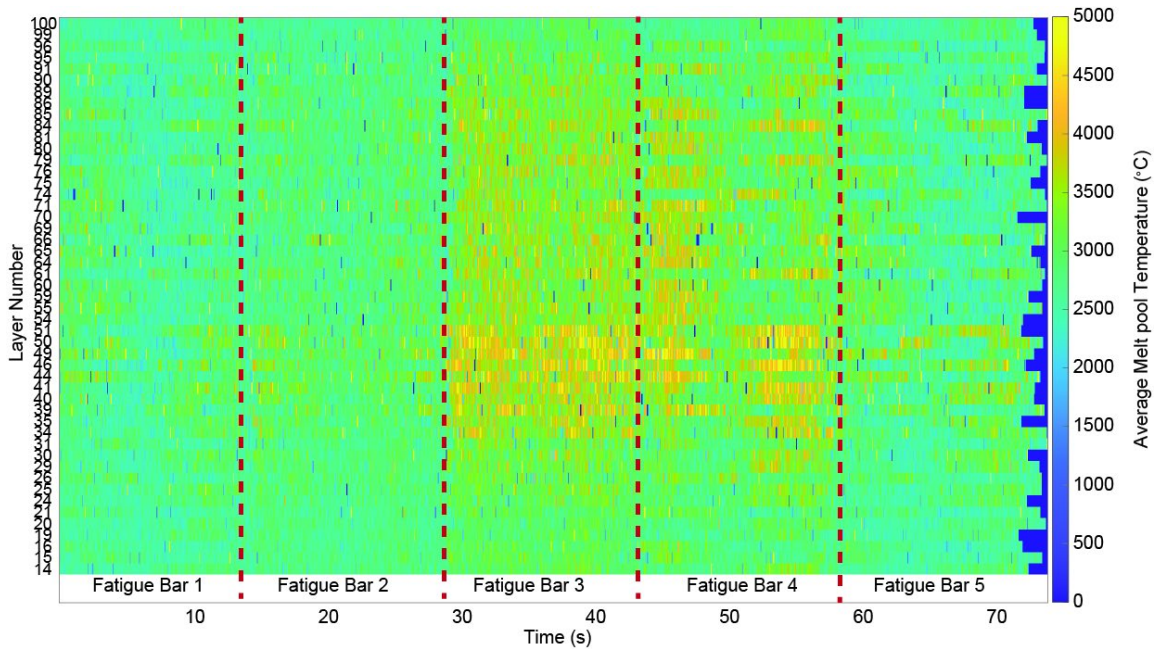

Figure 6 Average MP temperature of the five printed Fatigue Bars corresponding to 49 layers, derived from the MP intensity profiles acquired at the two working wavelengths of STWIP (550 nm and 620 nm). The red dashed lines show the approximate temporal boundaries of each printed Fatigue Bar. [please refer to web version of the article for color images]

## 4.2 Results and Discussions

The MP intensity profile for all the 49 layers corresponding to the two wavelengths (550nm and 620nm) were extracted as shown in Figure 5. Figure 5a represents the intensity profile of the five fatigue bars at 550nm, and Figure 5b shows the intensity profile at 620nm for these five fatigue bars. From these plots, the temporal distinction between the printed fatigue bars can be noticed (also marked with red dashed lines). This distinction represents the processing laser "off" time before proceeding to print the next bar, dividing the time sequence of MP data into five segments corresponding to each of the five labeled printed fatigue bars, respectively. Fatigue Bar 3 is approximately at the center of the build plate, Fatigue Bar 1 is closest to the bottom end of the build plate (closer to the machine door) and the Fatigue Bar 5 is at the top end of the end build plate (refer to Fig. 5a). It can be noted that the MP intensity for the first few layers (up to Layer 26) is higher compared to the subsequent Layers. And there is a sudden jump in the intensity profile from Layer 51 to 54, followed by a gradual increase in the intensity until the Final Layer (Layer 100). The extracted intensity profile is a combination of multiple factors occurring during the build. The ejected plume or vapors can affect the intensity profile [29], the surface conditions – such as recoating errors, exposure of previously printed layer can also affect the MP intensity profile. These observed changes in the intensity profile are possibly due to the surface conditions or the ejected vapors blocking the camera's optical path. It should also be noted that the intensity variation from bar to bar also varies in the same layer, for instance, the intensity values gradually increase from Bar 1 to Bar 3 and decrease Bar 3 to Bar 5 from Layers 54 through 100. This variation is more prominent in Figure 5a corresponding to the 550nm wavelength. This variation in the intensity profile could possibly be due the processing laser's intensity variation across the build plate and the location of the printed part on the build plate (refer Fig. 5a). The intermittent pixel specks in these plots are due to the camera sensor noise. For better visualization, these specks can be filtered using filtering approaches such as a moving average filter.

From these intensity profiles, we extracted the average MP temperatures for the 5 Fatigue Bars across all the acquired Layers, as shown in Figure 6. It can be clearly noted that the average temperature for the Bar 3 is higher than the rest of the Bars. The temperature rise is evidently seen at Layers 31 through 51. Based on this analysis, we conclude that the commonly used single wavelength intensity analysis without the actual emissivity information can lead to false inferences. And the mere analysis of the MP intensity profile can be incomplete and inaccurate. For instance, from the intensity profiles shown in Figs.5b and 5c, it can be wrongly presumed that the temperature profile of the MP can be linearly correlated to the intensity profile (by assuming a constant emissivity value for the material). But our result from the two-wavelength thermography method (Figure 6) shows evidently that the temperature profile is uniform without any abrupt fluctuations as opposed to what is indicated by the single-wavelength intensity profiles. Therefore, we conclude that a two-wavelength analysis such as the STWIP method can provide a comprehensive understanding of the temperature profile compared to emissivity dependent methods that rely on a single wavelength measurement.

Furthermore, we analyzed the MP temperature at the critical gage section of the fatigue bar 3 for Layers 14, 60, and 100. Fatigue bar 3 was chosen based on the observed disparate temperature profile, which might indicate potential process anomaly and or part defects (Fig. 6). For brevity, the 2D temperature maps of Layer 14 are shown here in the main text and the other layer's data in presented in the SI. The objective of this analysis was to understand the variation in the MP temperature profile for a critical portion across the layers during the process, evaluate the potential of using the measured MP temperatures as a representative signature for process-structure-property analysis and as a prospective online tool to offer insight of thermodynamics for LPBF process modeling and control. The schematic is Figure 7a shows the approximate locations corresponding to the temperature maps reported in the Figures 7 (b-d), S13 and S14. The color bar has the temperature limits of 0-5000°C. And the blue background in the images indicate the absence of MP pixels, thus indicating a pseudo zero temperature region. The temperature maps shown here are specifically focused to show the melt pool core, therefore the melt pool tail is not visible in these images. However, we do observe discernible melt pool tails of 500-600 μm through our STWIP system (SI Figure S15). Each Fatigue Bar print elapses ~14.5 seconds, and the gage section elapses approximately 4 seconds (the mid-portion of the print). Therefore, the frames corresponding to these 4 seconds at Layers 14, 60, and 100 were separately extracted for evaluating the temperature maps. From the extracted frames, the beginning, middle and ending frames, each set corresponding to approximately 20 milliseconds are further evaluated for determining their temperature profiles.

From Figures. 7b, S13a and S14a (beginning print of the gage section at Layer 14, 60, and 100, respectively), it can be noted that the MPs at the start time, have a relatively smaller width and lower temperatures. As the MPs progress temporally, we can observe the changes in the MP width/length temperature and scan direction. For instance, note the temperature changes in Figure 7b (beginning print of the Reduced region at Layer 14) from the start point to 4ms, there is a significant increase in the MP temperature followed by a cool down at 5ms and 10ms. From these temperature maps, it can also be seen that some MPs have very high temperatures at the edges, rather the center, see for instance MPs at (i) 4ms in Fig. 7b, (ii) 16ms in Fig. 7c, (iii) 0ms in Fig. 7d, (iv) 1ms in Fig. S13a and (v) 2ms in Fig. S14c. These high temperatures could possibly be due to (i) incorrect image transformation (KAZE) or (ii) some process defect causing high temperatures. We are currently working on improving the accuracy of our image transformation algorithm and, integrating an independent off-axis camera and ex-situ analysis of that location can possibly help notice this irregularity in the MP temperature.

Also, note the directionality of the MP pairs at 3 and 4, 6 and 7, 9 and 10, 12 and 13 milliseconds in Figure 7b, from this data, we can understand the directionality of the MP's travel and by closely analyzing the MP intensity profile we could possibly evaluate the hatching pattern corresponding to each layer. This information can be helpful when developing machine learning frameworks for determining the print and process errors. However, with an additional data signature from an independent off-axis camera, which captures the print process covering the whole build plate, the exact position of the MP can be registered, and a comprehensive MP data signature can be evaluated.

Another key observation to be noted from these temperature profiles is the relative position of the MP at different time stamps. The start, middle, and end point of Reduced region was chosen purely based on the theoretical estimation of the frame number and no knowledge of the print direction/hatch pattern. It can be noted from Figure 7b, S14a (beginning print of the gage section at Layer 14, and 100) and Figure 7d, and S14c (ending print of the gage section at Layer 14, and 100) that the MP center shift from px # 38 to px # 41 – clearly showing a shifting in the MP position across the build plate mainly due to the position shift in the F-$\theta$ lens. However, the trend for Layer 60 (Figure S13a – S13c)) is different. The MP center for this Layer moves from px # 41 to px # 38 indicating a different print direction than Layers 14 and 100. It has to be noted that 1 pixel is equal to 20 µm, using this scale, we observe a shift of at least 60 µm. This shift is probably due to a different hatch pattern corresponding to this Layer. From these Figures (7, S13 and S14), we can also estimate the MP width and area. Currently, we are performing independent experiments to validate the MP width measured using our in-situ method with ex-situ characterization methods such as SEM. Preliminary analysis showed that our measurements are within the 5% limits of the MP widths measured using ex-situ characterization methods. Further, the changes observed in the average temperature profile of the Fatigue Bars 3 and 4 (Fig. 6) will be validated with the ex-situ measurements (such as microscopy, X-ray CT, mechanical tests) to understand the observed change.

Although, the 2D temperature plots presented here are preliminary and only correspond to few specific locations on the fatigue samples, this analysis paves way for a deeper understanding of the melt pool temperature variation at the desired part location for a practical build case. Further development of the feature recognition and transformation algorithms along with parallelization of the image transformation code is required for the large scale analysis and comparison with the ex-situ measurements. We are currently in the process of improving these algorithms for a 2D temperature better analysis.

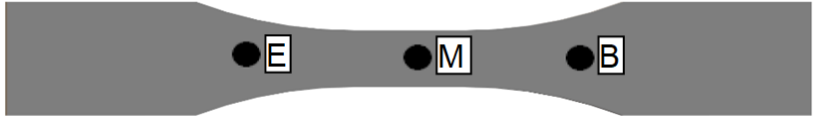

(a)

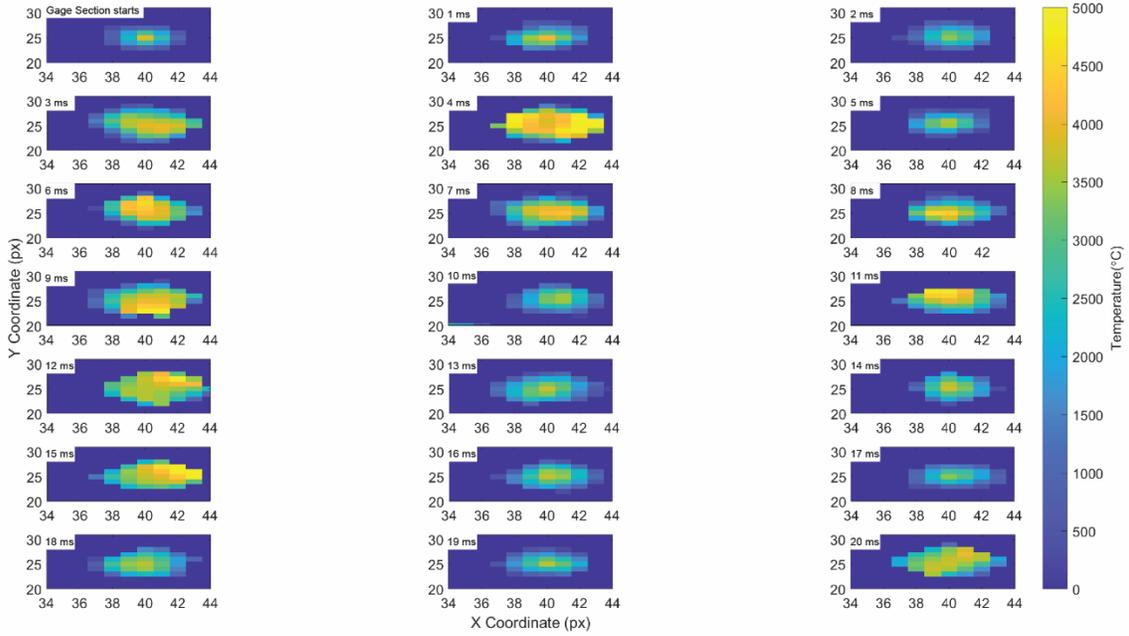

(b)

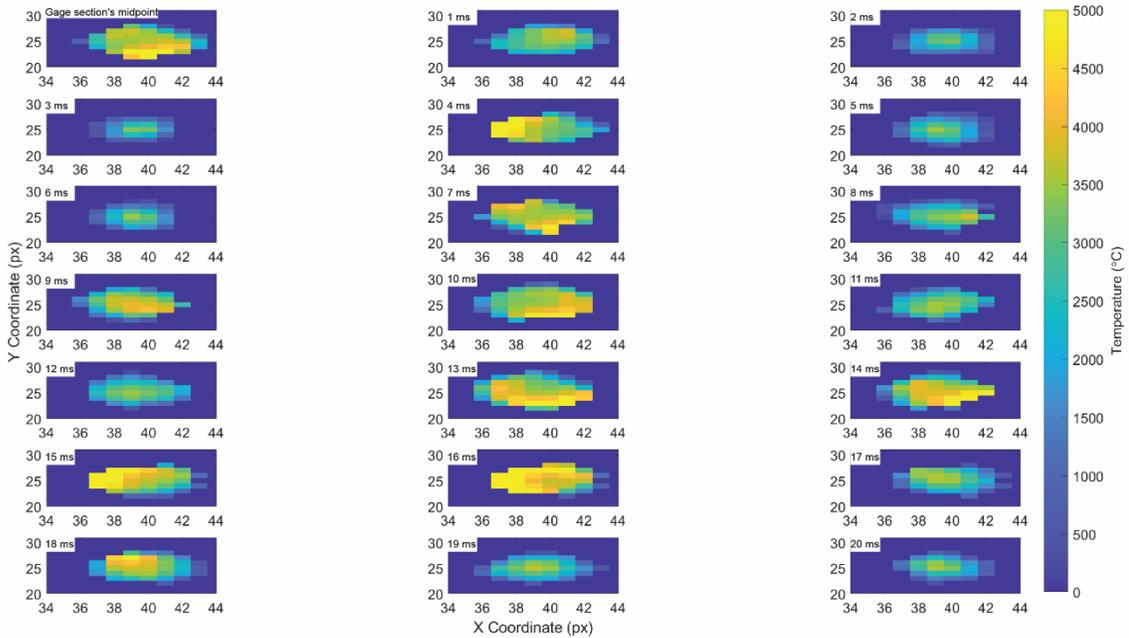

(c)

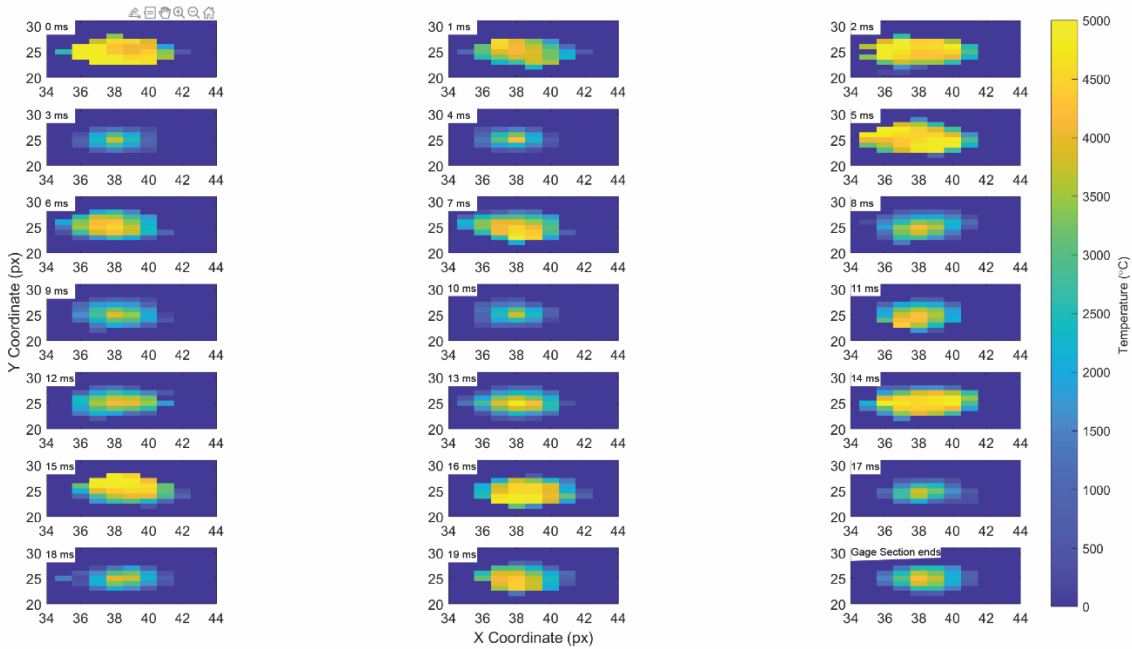

(d)

Figure 7 (a) Schematic showing the melt pool core's temperature measurement locations. B indicates Beginning, M indicates Middle portion and E indicates the End portion of the gauge Section. (b) Beginning, (c) Middle and (d) End gage section's melt pool core's temperature profile measured by the STWIP for Layer 14 in a test sample (Fatigue Bar 3), temporally separated by 1ms. Note: the blue pixels surrounding the detected MP boundary are pseudo background pixels and do not necessarily mean $0°C$. [please refer to web version of the article for color images].

## 5 Conclusion and Recommendation

In this work, we have demonstrated that the developed Single-camera Two-Wavelength Imaging Pyrometry (STWIP) system and methods can successfully estimate the 2D temperature profile of the melt pools and their morphology, as well as track their progression and variation during LPBF metal AM process. The performance of STWIP is evaluated through the following investigations. (1) A facile experimental validation against thermocouple measurements to assess its accuracy (>90% for all the tested cases and projected to be >95% for a real test case) and precision (repeatability >95% for all the tested cases and projected to be >98% for a real test case), which can be further validated to exhibit even better performance given a more capable validation approach and experimental setup. (2) A quantitative uncertainty analysis of temperature measurement by incorporating the impacts of the optical path efficiency and the camera sensor(relative uncertainty <2.8%; and absolute uncertainty is, for example, 31.71°C for a measured temperature of 2760°C corresponding to a typical Intensity ratio of 1.1 as observed in the LPBF process. (3) An application case of implementing STWIP for MP intensity profile and temperature measurement during a print of five fatigue test samples to assess the advantage, capability, and usage of STWIP. The elucidated uncertainty analysis offers quantitative insights about the capability and potential of STWIP in MP temperature measurement, enhancing the confidence and utility of STWIP for the AM community to perform research on LPBF and other metal AM processes. The MP temperature measurement and monitoring is successfully demonstrated for multiple layers in a representative and practical test case. Although single sensor based two-wavelength intensity pyrometry is not new, to the best of our knowledge, we are the first to report a study that demonstrates explicitly the feasibility, capability, and potential of employing a coaxial STWIP for MP monitoring for a practical application case in LPBF process monitoring at fast rates, larger data sets, and monitoring scales (30,000 fps, ~ 2.2 million images or 45 GB of image data (raw + segmented) per layer for 49 layers of 5 fatigue bars, in total of ~1 billion melt pools analyzed and measured). The STWIP method significantly enhances the capability of high-speed large-scale data acquisition, transfer, and analytics, enables a continuous MP monitoring for much longer time, and heightens the MP temperature measurement efficiency without sacrificing accuracy or resolutions (both temporal and spatial).

Moving forward, the algorithms of two-wavelength MP transformation and temperature estimation will be further improved for more accurate and rapid processing to perform online data analytics and temperature calculations for achieving real-time MP measurement and feedback control of LPBF. Given more capable hardware and faster computational platform, the frame rates, thus the MP monitoring speed could be further increased as per the user's utility. Registration of MP data will be accomplished by integrating the developed coaxial system with an off-axis high speed camera to track the MP on the physical part for subsequent process-property correlation analysis. Further, we will also validate the emissivity ratio ($\varepsilon_1 : \varepsilon_2$) for different materials through independent temperature measurement validations and perform an uncertainty analysis for understanding the changes in the MP temperature due to the emissivity ratio.

Furthermore, an extensive STWIP dataset of MP temperature signatures, such as each MP's average / peak temperatures and 2D temperature field, for various processing conditions will be built. A machine learning framework will be developed to correlate the derived spatiotemporal representation of MP temperature signatures to the printed part properties such as porosity, grain size and morphology, and various mechanical strength metrics.

Other than monitoring MP temperature, the developed STWIP system can directly monitor MP morphology such as MP width, shape and area. It is also worth noting that STWIP can be readily extended to become a multiple-wavelength imaging pyrometer or multispectral imaging based spectrometer for enhanced accuracy in temperature measurement as well as expanded capability of monitoring temperature and other key indicators such as phase transformation and elements [38, 39], by adding additional wavelength arms without requiring additional cameras or hindering measurement speed and duration.

To conclude, the evident capability and remarkable potential of STWIP will allow for affordable, efficient, robust, and continuous monitoring of LPBF and other metal AM processes such as wire-arc AM [40], to provide accessible (for broader research impacts and users) and enhanced insights into the energy beam and metal interactions and process thermodynamics.. Notably, the STWIP methodology can be extended to be a multispectral imaging pyrometer or spectrometer for more accurate temperature measurement and other process signatures monitoring. Harnessing advanced data science and computational technologies, the developed STWIP methodology holds great promise in facilitating the development of process models, the validation of simulations, the realization of real-time precision measurement and control, as well as achieve rapid cost-effective qualification of metal AM for broad industry applications, for enhancing the performance and utility of not only LPBF but also other AM processes. Besides, the STWIP and its variant systems can also be used in other research areas such as combustion related research in aerospace discipline.

# 6 Acknowledgements

**Funding:** Authors acknowledge the funding from the Department of Energy University Turbine Systems Research grant (DE-FOA-0001993 Award Number: FE0031774), ANSYS Additive Manufacturing Research Laboratory (AMRL) at University of Pittsburgh.

Authors would also like to thank Brandon Blasko, Yubo Xiong and Dian Li for their help with the experiments. This work used the Extreme Science and Engineering Discovery Environment (XSEDE) resource - Pittsburgh Supercomputing Center Bridges GPU and Storage through allocation MCH200015.

# 7 Declaration of Competing Interest

Some of the results and remarks about the development and application of STWIP are part of a pending patent (USPTO Application No.: 17/015,062, filed on September 8, 2020. Priority Date: November 8, 2019).

# 8 CRediT Author Statement

**Chaitanya Krishna Prasad Vallabh**: Methodology, Software, Investigation (Experiments and Data acquisition), Data Analysis, Writing-Original draft preparation & Editing, Visualization.

**Xiayun Zhao**: Conceptualization, Methodology, Resources, Writing – Draft, Review & Editing, Supervision, Project administration, and Funding acquisition.

# Supplemental Information

**SI Section I: Representative spectrum of melt pool radiance during LPBF of IN718.**

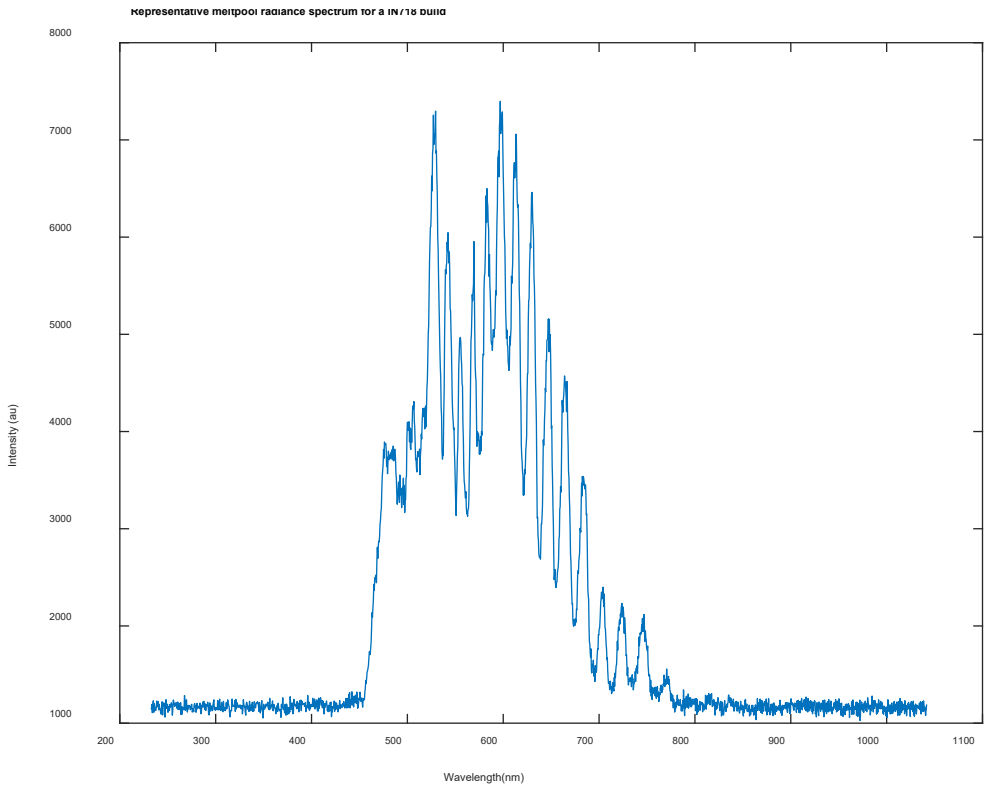

Figure S1: Representative spectral response from LBPF of Inconel IN718 Powder

**SI Section II: Optical calibration calculation details**

To eliminate this uncertainty in our optical path efficiency measurements, we performed $A_1/A_2$ (or $A_{12}$) ratio experiments using the experimental setup shown in Figure S2a. The $A_{12}$ value was characterized at five different locations on the build plate (Figure S2b) to understand the effect of the location on the $A_{12}$ ratio.

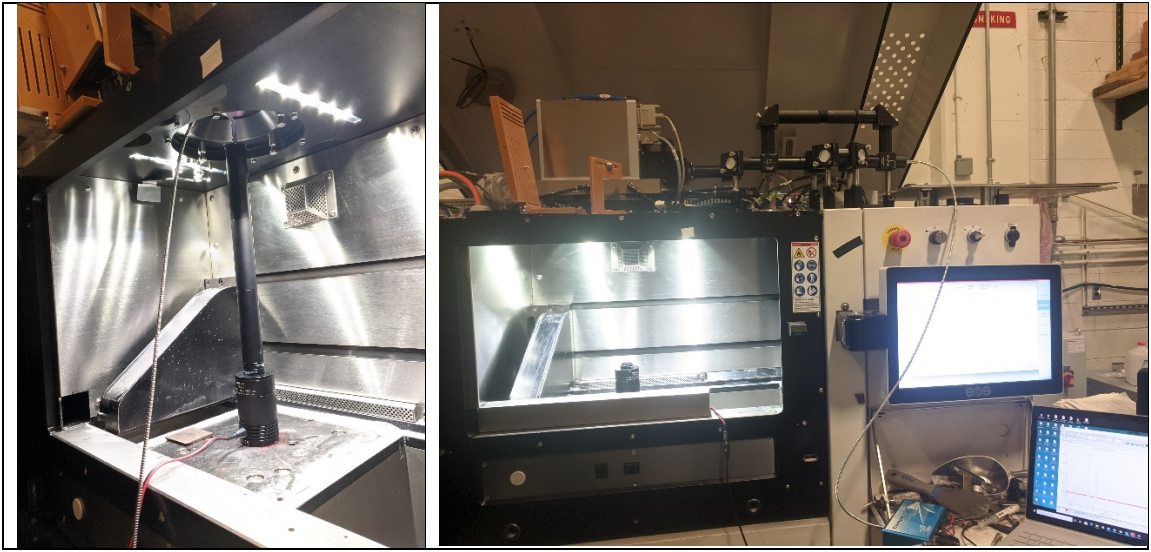

(a)

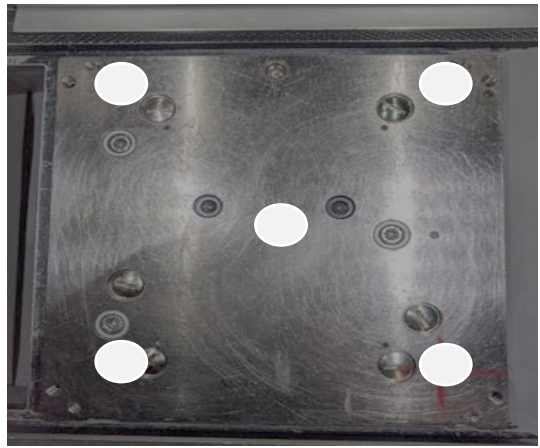

(b)

Figure S2. Experimental setup to characterize the $A_1/A_2$ ratio ($A_{12}$). (a) A light bulb is placed on the center of build plate with a neutral density (ND) filter mounted above. A tube lens guides the light source beam up towards the machine's F-θ lens and rigidly supports a spectrometer to ensure a stable reading. The spectrometer measures the light beam spectral intensity that is input to the F-θ lens. (Right) The spectrometer measures the output light beam at the outlet conduit of the developed in-situ monitoring system. This developed setup characterizes the optical path efficiency at five locations on the build plate as shown in (b), this figure shows the printer's build plate base and the approximate testing locations for evaluating the $A_{12}$. This validation helps determine the value of $A_{12}$ for robust and accurate temperature measurements.

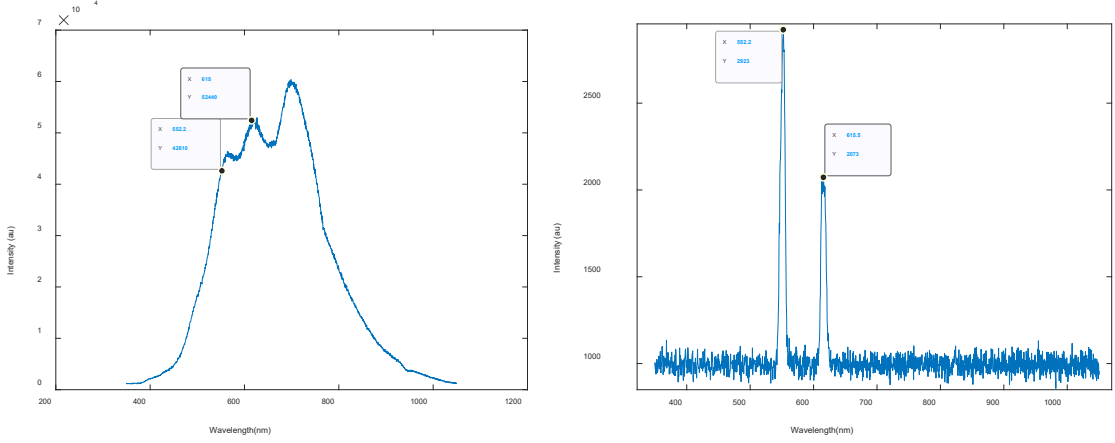

Figure S3: Representative Spectrometer's response of the input (light source) and output (optical path) measured as shown in Figure S2a.

The spectral transmission of the entire system of the commercial LPBF laser box and the lab-designed in-situ monitoring system (i.e., from the F-θ lens in the 3D printing machine to the end of the in-situ monitoring optics excluding the pre-camera focal lens and camera), $T_\lambda^{Sys}$ can be written as:

$$T_\lambda^{Sys} = \frac{I_\lambda^{Spectrometer\_Outlet}}{I_\lambda^{Spectrometer\_Inlet}}$$

Where $I_\lambda^{Spectrometer\_Outlet}$ and $I_\lambda^{Spectrometer\_Inlet}$ refer to the spectrometer readings at the end of the optical and at the light source, as shown in Figure S2a. Specifically, for wavelengths $\lambda_1$ (550nm) and $\lambda_2$ (620 nm), using above equation, we have:

$$T_{\lambda 1}^{Sys} = \frac{I_{\lambda 1}^{Spectrometer\_Outlet}}{I_{\lambda 1}^{Spectrometer\_Inlet}}$$

$$T_{\lambda 2}^{Sys} = \frac{I_{\lambda 2}^{Spectrometer\_Outlet}}{I_{\lambda 2}^{Spectrometer\_Inlet}}$$

The spectrometer's responses typically resemble the responses shown in the representative plots in Figure S3. Considering the transmission coefficients of the focus lens and the spectral response of the camera's sensor, the final equation can be written as:

$$A_{12} = \frac{T_{\lambda 1}^{FLen} \times R_{\lambda 1}^{Camera}}{T_{\lambda 2}^{FLen} \times R_{\lambda 2}^{Camera}} \times \frac{I_{\lambda 1}^{Spectrometer\_Outlet}/I_{\lambda 2}^{Spectrometer\_Outlet}}{I_{\lambda 1}^{Spectrometer\_Inlet}/I_{\lambda 2}^{Spectrometer\_Inlet}}$$

The transmission coefficients ($T_1^{FLen}$ and $T_2^{FLen}$) of the lens were determined from the manufacturer's specification sheet to be 0.98 and 1 for 550 nm and 620 nm, respectively. And the $R_1$ (0.85 for 550nm) and $R_2$ (0.98 for 620nm) camera's spectral response values for a monochrome sensor, which were estimated from the graph below (Figure S4).

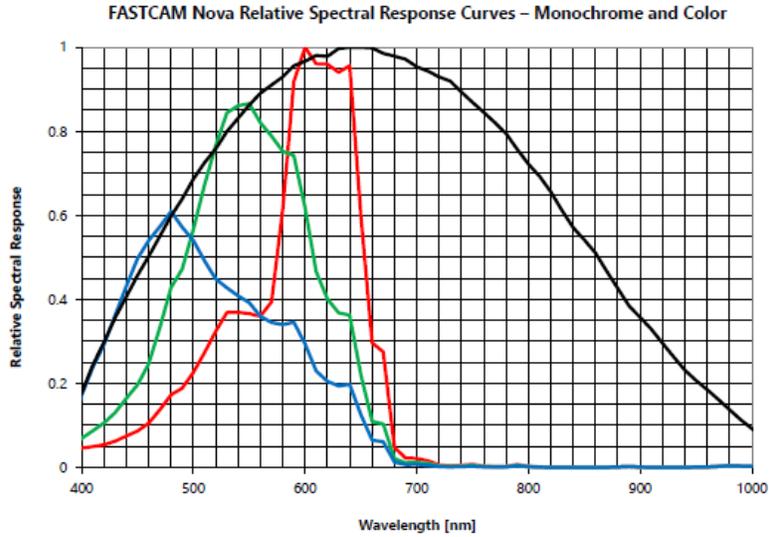

Figure S4: Spectral response graph of the Camera's Sensor. Our camera uses a monochrome sensor

Table S1: $A_{12}$ for the develop optical system validated at five different location on the build plate

|  | $A_{12}$ - Average values from 10 individual measurements at each location |
|---|---|
| **Center** | **1.608** |
| **Bottom right Corner** | **1.610** |
| **Top Left Corner** | **1.622** |
| **Bottom left Corner** | **1.579** |
| **Top Right Corner** | **1.585** |
| **Average** | **1.601** |
| **Standard Deviation** | **0.0163** |

From these experiments, we characterized $A_{12}$ value accurately, with a very low measurement error (Table S1). This observation was further validated by a statistical analysis. The ANOVA statistical analysis for the $A_{12}$ data showed that the location has no significant effect on the $A_{12}$ ratio. ANOVA tests the hypothesis that all group means are equal versus the alternative hypothesis that at least one group is different from the others. From our analysis, we do not reject the Null Hypothesis; that is the mean values of all the groups are similar to each other as shown in the figure below (Figure S5). Ten individual measurements were acquired at each location on the build plate and compared for evaluating the final $A_{12}$ ratio of our system. The final $A_{12}$ ratio of our system is 1.601±0.01063.

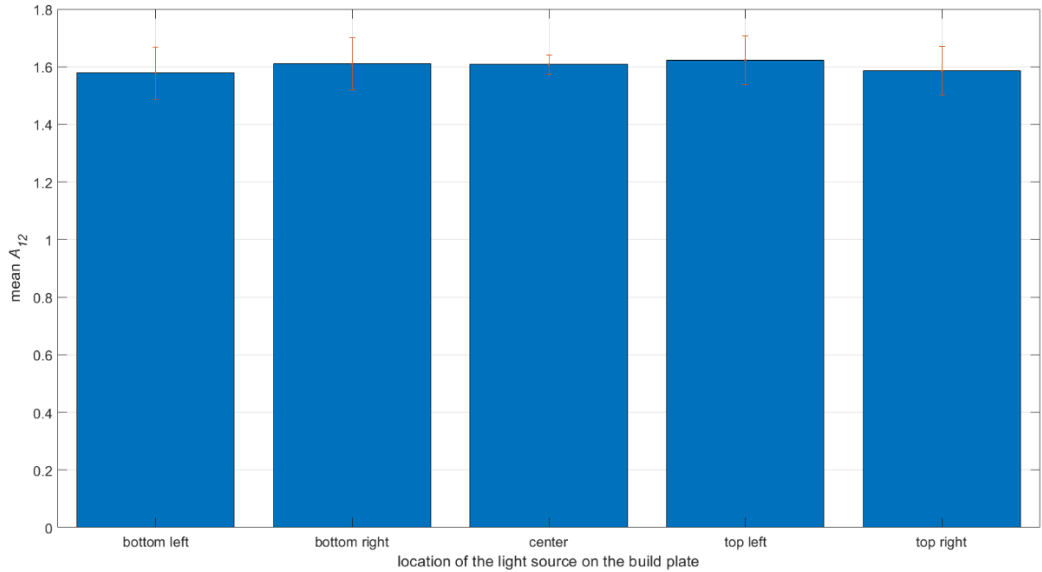

Figure S5. Bar graph representation of the mean A1/A2 ratios for different locations on the build plate

**SI Section III: Verifying the two-wavelength image transformation**

Validation of image transformation was performed using – Structural similarity (SSIM) index for measuring image quality in MATLAB. The SSIM Index quality assessment index is based on the computation of three terms, namely the luminance term, the contrast term and the structural term. The overall index is a multiplicative combination of the three terms. The SSIM Index quality assessment index is based on the computation of three terms, namely the luminance term, the contrast term and the structural term. The overall index is a multiplicative combination of the three terms.

SSIM $(x,y)=$[luminance $(x,y)]^{\alpha} \cdot$[contrast $(x,y)]^{\beta} \cdot$[structural $(x,y)]^{\gamma}$, where $\alpha$, $\beta$, and $\gamma$ are the respective exponents of luminance, contrast and structural terms. X and Y are the image's spatial pixel information.

Please refer to MATLAB's documentation on SSIM for the details on the algorithm (https://www.mathworks.com/help/images/ref/ssim.html).

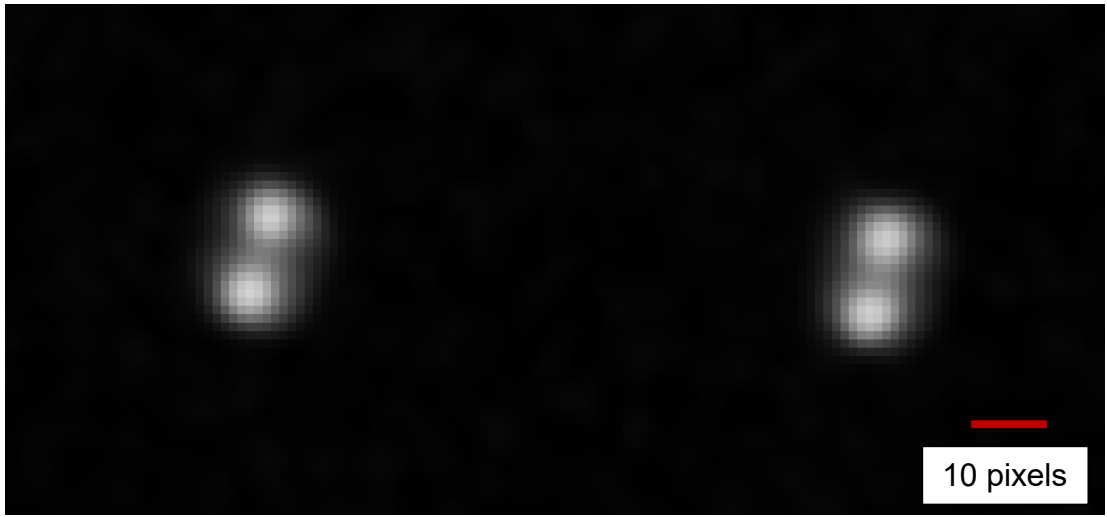

(a)

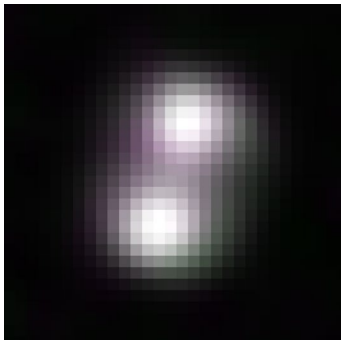

(b)

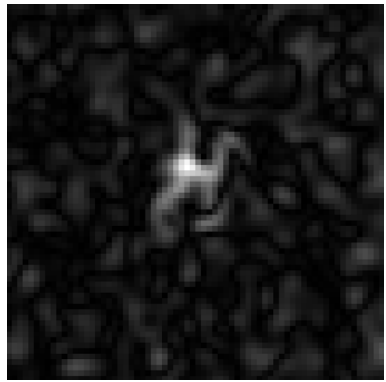

(c)

Figure S6: Verifying the KAZE feature recognition algortihm with a known image shape. (a) Original two wavelength images. Image on the right is transformed with reference to the image on the left. (b) Image transformed using KAZE and superimposed on the reference image and (c) Difference between the original and the tranformed image. The SSIM index for these two images was 0.9663, indicating a very high matchness/similarity of the transformed image to the reference image, thus validating the employed KAZE algorithm.

# SI Section IV: Experimental validation of STWIP

The temperature measurements of our in-situ STWIP system were experimentally validated using a thermocouple experimental test case. The experimental setup is shown in Figure S7a. As shown in the Figure S7a, the thermocouples are embedded inside a 4x4 inch Inconel build plate. Two different scanning patterns thus two test cases (Figure S7b) were used for these validation experiments.. The schematics of the scans are shown in Fig. S7b. For test case #1 we used a simple rectangle pattern (10×20 mm) and for test case #2 we used a star-like pattern in a rectangular array of 10×20 mm as shown in the schematics. Test case #2 measurements matched closely to the in-situ measurements with lower deviation values as shown in Table S2. This could be due to the employed design, where in the scanned pattern could have better interacted with thermocouple sensing tip compared to the pattern in test case #1, resulting in the observed variations in Table S2. The scan patterns for test case # 1 were laser power of 25W and scan speed of 150mm/s.And for test case #2 laser power of 50W and scan speed 50 mm/s were used.

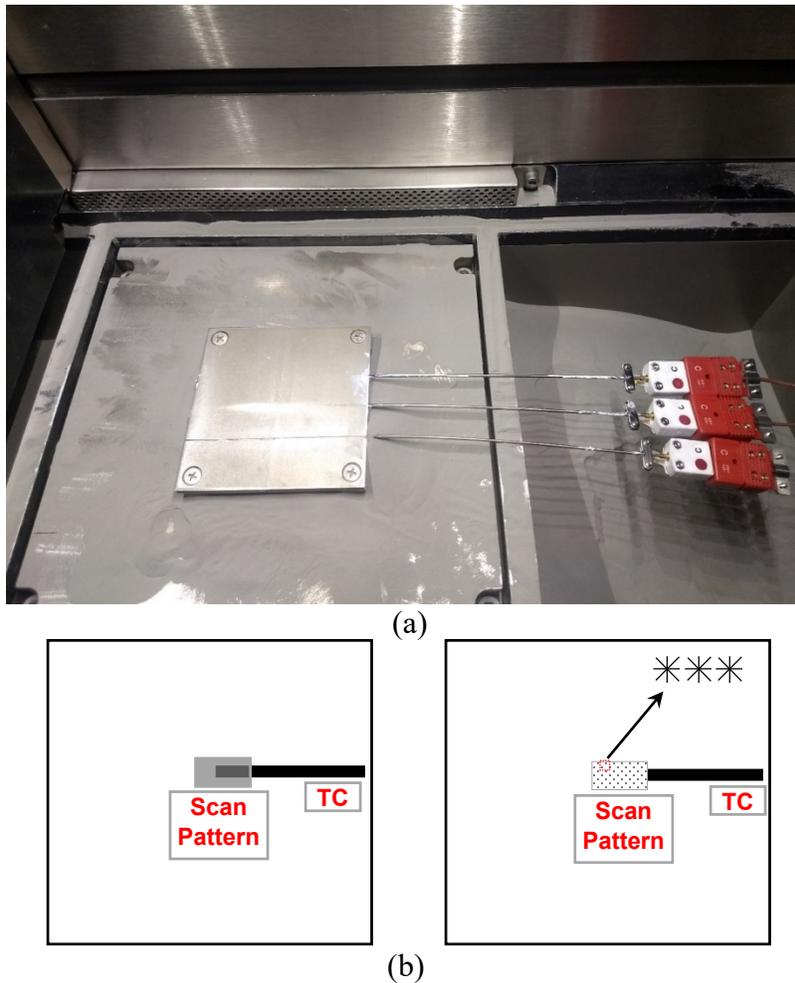

Figure S7: (a) Thermocouple setup for the temperature validation experiments. The thermocouples are placed inside the build plate, approximately 80 μm deep from the surface of the plate. At this depth, the thermocouples can detect the MP thereby estimating their relative temperature. (b) scan schematics used for Test Case #1 - 10×20mm rectangle (left) and Test Case #2 10×20mm rectangle with star-like patterns as shown in the figure on the right. The star-like

patterns have a diameter of 3mm (not-to-scale). The box represents the laser scanning pattern and the dark solid line indicates the thermocouple (TC) location.

The thermocouples used in our experiments have a time constant of ~0.33 seconds. For a thermocouple to register a 100% temperature change, the tip must be exposed at least for 5 time constants, i.e., 5×0.33s = 1.65s. Due to the EOS system limitations we could only expose the thermocuple (under the MP) for less than 0.5 seconds. Due to this limitation, the voltage responses of the thermocouples in the experiments correspond to ~63.2 % of the temperature. Based on this, we can estimate the actual MP temperature and compare it to the in-situ STWIP measurement results. For instance, one data set recorded a voltage of 31.92 mV, corresponding to 1862°C (63.2% of the temperature change) at the time point of ~33.18 s as shown in Figure S8a, which results in the actual temperature to be 2947.92°C. The evaluated MP temperature (2947.92 °C) is considered to be the average MP temperature, as the thermocouple's element diameter (~3.175 mm) is much larger than MP width and length (typically ~100's of µm). The MP temperature for the same scan estimated from our STWIP system, 3311.82 °C (Fig. S8(c)), is close to the average MP temperature estimated from the thermocouples (2947.92 °C).

The STWIP estimated temperature, at same time stamp (~33.18 s) corresponding to the thermocouple measurement, is~3534 °C. This high value might be due to the sudden increase in the MP intensity when laser interacts with the build plate at the thermocouple location, as observed in the intensity profile plot (Fig. S8 (b))..

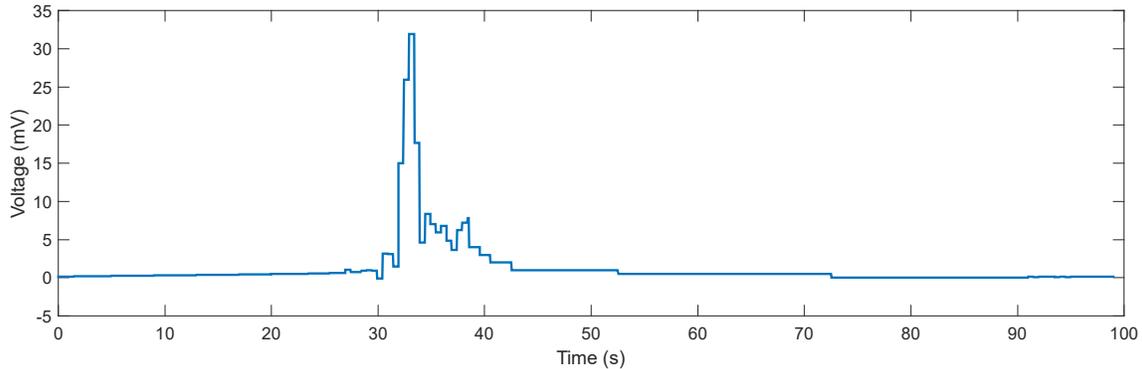

(a)

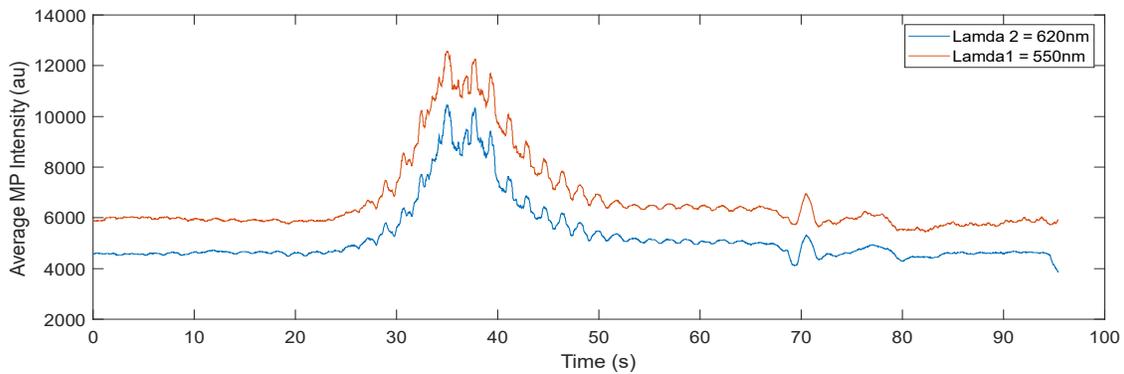

(b)

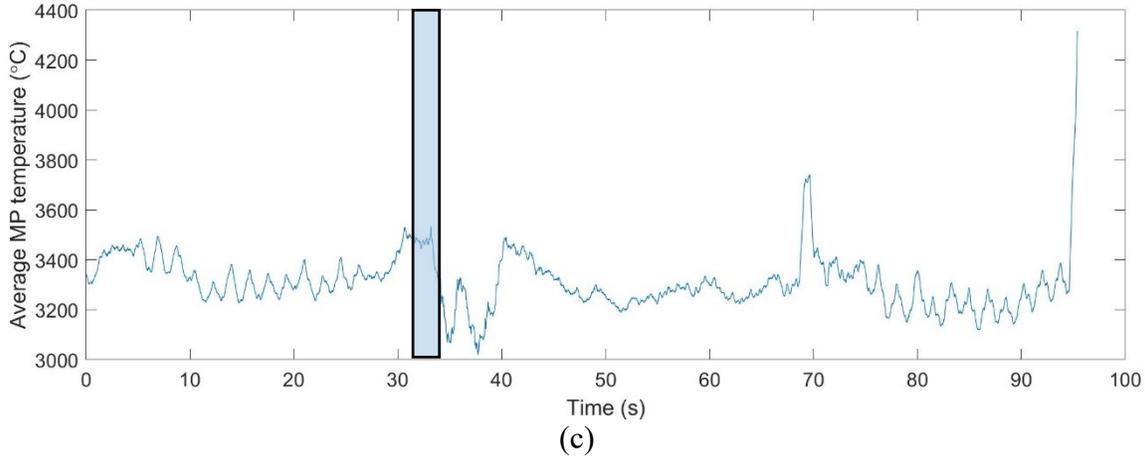

(c)

Figure S8 (a) Representative thermocouple voltage measurement corresponding to the temperature validation experiments. The peak in the graph represents the temperature step change in the MP near the thermocouple tip. (b) MP average intensity plots and the (c) average MP temperature plot for the thermocouple temperature validation experiment shown in (a). The intensity peak in the plot (b) shows the increase in the intensity when the laser interacts with thermocouple inside the build plate (Figure S7). The shaded rectangle in the plot corresponds to the thermocouple response of 31.92mV (~33.18s).

The graphs (Fig. S8) show the thermocouple data corresponding to one thermocouple measurement, the MP intensity, and the average MP temperature derived from our in-situ system for the same experiment. Five individual thermocouple measurement experiments were performed for evaluating the accuracy and repeatability of our in-situ system. Measurement results for one thermocouple (top thermocouple in the schematic Fig. S7) are shown in Table S2. The accuracy of the STWIP system can be further validated using other established methods, such as calibrated lamps, validation against infrared temperature cameras.

Table S2: Estimated MP temperature from the thermocouple measurements along with the corresponding in-situ coaxial system estimated average MP temperatures. The percentage difference between the cumulative measurements was found to be 8.16 ± 5.16%. 1.1 – 1.3 correspond to Test case 1 and 2.1 and 2.2 correspond to test case 2. Test case 1 uses a large rectangular scan pattern and test case 2 uses a small star-like patterns (ref. Fig S7b)

| Test Case | MP temperature estimated from Thermocouple Responses – $T_t$(°C) | Average MP temperature from Co-axial system Response – $T_c$ (°C) | Relative Measurement Difference $|((T_c - T_t)/T_t)| \times 100\%$ |
|---|---|---|---|
| 1.1 | 3247.71 | 3726.003 | 14.72% |
| 1.2 | 2947.92 | 3311.82 | 12.34% |
| 1.3 | 3382.95 | 3694.18 | 9.19% |
| | | Test Case #1: Average ± SD | 12.08%±2.77% |
| 2.1 | 3091.98 | 3053.89. | 1.23% |
| 2.2 | 3797.52 | 3670.12 | 3.35% |
| | | Test Case #2: Average ± SD | 2.29%±1.50% |
| | | Overall (two cases) Average ± SD | 8.16%±5.76% |

**SI Section V: Temperature Measurement Uncertainty Analysis**

Detailed analysis of temperature measurement uncertainty is conducted as below based on Equations (3) and (4).

For illustration, the temperature measurement uncertainty due to $A_{12}$ is exemplified as below. Here, we use a typical value as observed in the experiments, i.e., $I_{12} = 1.1$ (see Fig. S11). $U_{I_{12}}$ and $U_{\lambda_{12}}$ are neglected for this analysis. For our STWIP system, $A_{12} = 1.601 \pm 0.0163$, thus $U_{A_{12}} = 0.0163$. $\theta_{A_{12}}$ is the sensitivity of $A_{12}$. $h = 6.626 \times 10^{-34}$; $k_b = 1.380649 \times 10^{-23}$; $c = 3 \times 10^8$ m/s; $\lambda_1 = 550$nm; $\lambda_2 = 620$nm.

$$T = \frac{\frac{hc}{k_B}(\frac{1}{\lambda_2} - \frac{1}{\lambda_1})}{\ln(I_{12}) - 5\ln\left(\frac{\lambda_2}{\lambda_1}\right) - \ln(A_{12})}$$

$$U_{T_{A_{12}}} = \sqrt{\left(\theta_{A_{12}} U_{A_{12}}\right)^2}$$

$$U_{T_{A_{12}}} = \sqrt{\left(-2955 * \left(\frac{\partial}{\partial A_{12}}\left(\frac{1}{\ln(I_{12}) + -5\ln\left(\frac{\lambda_2}{\lambda_1}\right) - \ln(A_{12})}\right)\right) U_{A_{12}}\right)^2}$$

$$U_{T_{A_{12}}} = \sqrt{\left(-2955.5 * \left(\frac{1}{A_{12}\left(\ln(I_{12}) - 5\ln\left(\frac{\lambda_2}{\lambda_1}\right) - \ln(A_{12})\right)^2}\right) U_{A_{12}}\right)^2}$$

$$U_{T_{A_{12}}} = \sqrt{\left(-2955.5 * \left(\frac{1}{1.601(0.0953 - 0.5990 - \ln(1.601))^2}\right) 0.016\right)^2}$$

$$U_{T_{A_{12}}} = \pm 31.71°C$$

Similarly, the temperature measurement uncertainty due to $I_{12}$ is exemplified as below. Using the known $A_{12} = 1.601$; $U_{A_{12}}$ and $U_{\lambda_{12}}$ are neglected for this analysis. Using the same typical value as above, i.e., $I_{12} = 1.1$ and $U_{I_{12}} = 0.003$. $\theta_{I_{12}}$ is the sensitivity of $I_{12}$. $h = 6.626 \times 10^{-34}$; $k_b = 1.380649 \times 10^{-23}$; $c = 3 \times 10^8$ m/s; $\lambda_1 = 550$nm; $\lambda_2 = 620$nm.

$$U_{T_{I_{12}}} = \sqrt{\left(-2955.5 * \left(\frac{1}{1.1(0.0953 - 0.5990 - \ln(1.601))^2}\right) 0.0003\right)^2}$$

$$U_{T_{I_{12}}} = 0.849 \,°C$$

The $U_T$ due to $A_{12}$ and $I_{12}$ are shown in the plots below. From the plots, it is evident that the uncertainty due to $I_{12}$ is minimal and the uncertainty due $A_{12}$ decreases as the $A_{12}$ value increases.

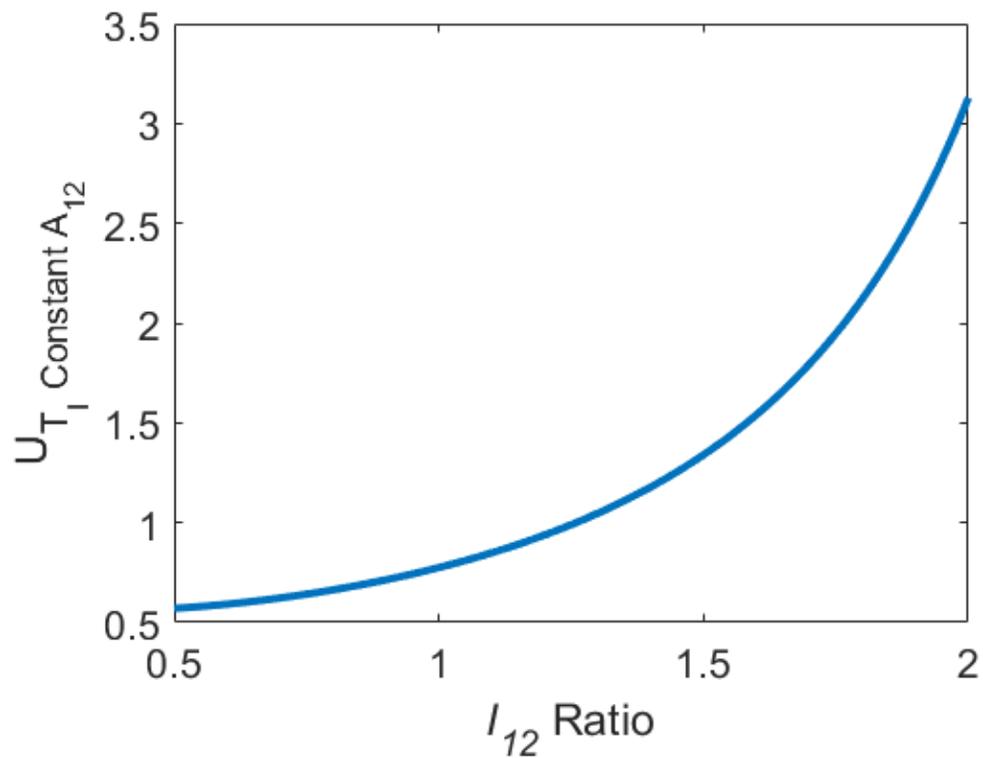

(a)

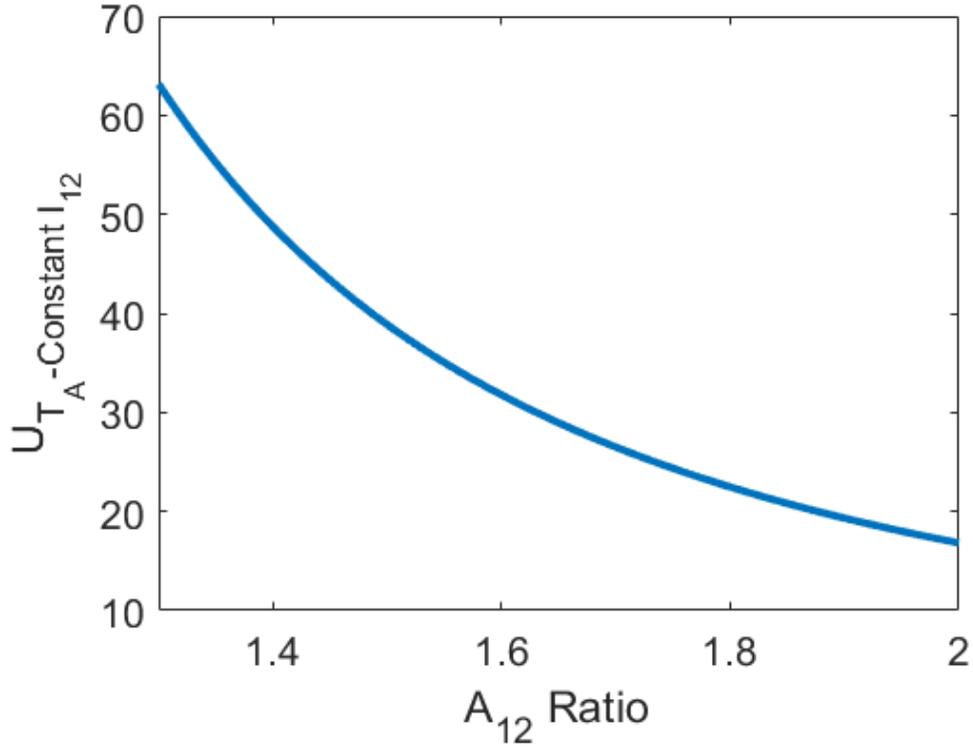

(b)

Figure S9: Uncertainty in Temperature due to (a) $I_{12}$ with an uncertainty of $U_{I_{12}} = 0.0003$ for different a range $I_{12}$ values (0.5-2) for a constant $A_{12} = 1.601$ and (b) $A_{12}$ with an uncertainty of $U_{A_{12}} = 0.0163$ for a range of $A_{12}$ values (1.3 - 2) for a representative value $I_{12}=1.1$ .

Combining these two uncertainties due A12 and I12, we get

$$U_{T_{A-I_{12}}} = \sqrt{(\theta_{A_{12}} U_{A_{12}})^2 + (\theta_{I_{12}} U_{I_{12}})^2}$$

$$U_{T_{A-I_{12}}} = \sqrt{(U_{T_{A_{12}}})^2 + (U_{T_{I_{12}}})^2}$$

From the above relation, we evaluate the uncertainty in temperature due to both $A_{12}$ and $I_{12}$. The plot below shows the uncertainty in temperature vs the actual temperature for a constant $A_{12}$ value of 1.601 and for varying $I_{12}$ values of 0.5-2. From this plot, it can be concluded that the relative uncertainty in temperature evaluated from our system is less than 2.801%. This relative uncertainty (2.801%) is the maximum relative uncertainty of the system evaluated at $I_{12} = 2$. An $I_{12}$ ratio of 2 is very uncommon is typical print case scenarios. For a typical $I_{12}$ value of 1.1, the temperature of the MP evaluated from STWIP is 2760°C, for this temperature, the evaluated uncertainty is 31.71°C (Fig. S10). Therefore, the measured temperature for $I_{12} = 1.1$ is $2760 \pm 31.71$°C.

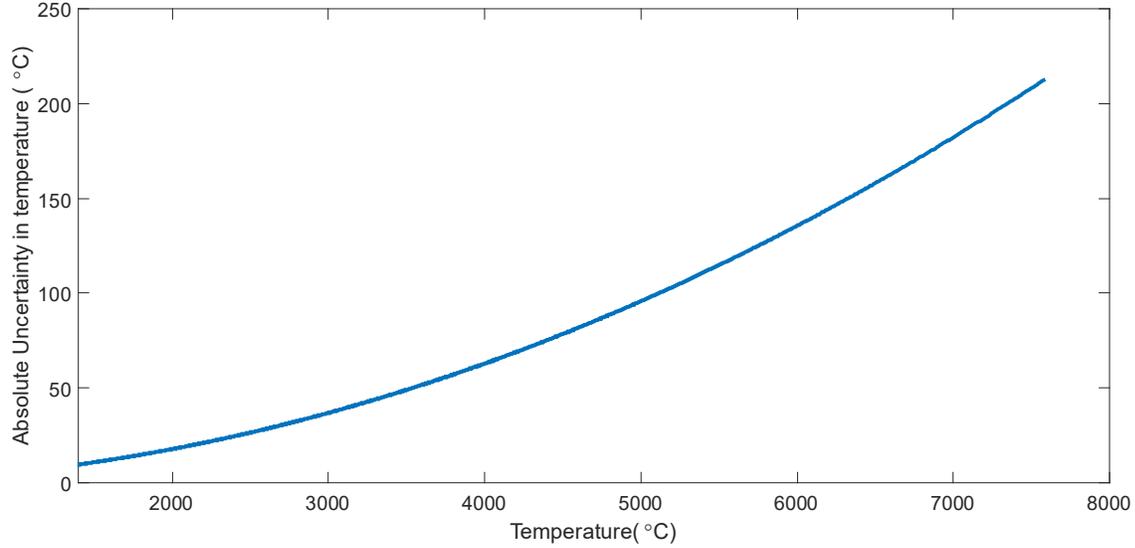

Figure S10: $U_T$ vs $T$ curve for $I_{12}$ varying from 0.5 to 2 for a $A_{12}$ value of 1.601

**Evaluating the $U_{I12}$**

For evaluating the uncertainty in $I_{12}$, we take the camera sensor's uncertainty into account. The smallest intensity increment of the camera sensor is 1 unit, therefore the uncertainty is $\pm 1/2$ unit. Using this relation, we evaluate the $U_{I12}$ as shown below:

$$U_{I_{12}} = \sqrt{\left(\frac{\partial I_{12}}{\partial I_1} U_{I_1}\right)^2 + \left(\frac{\partial I_{12}}{\partial I_2} U_{I_2}\right)^2}$$

where, $I_{12} = \frac{I_1}{I_2}$

$$U_{I_{12}} = \sqrt{\left(\frac{1}{I_2} U_{I_1}\right)^2 + \left(-\frac{I_1}{I_2^2} U_{I_2}\right)^2}$$

For typical Intensity values of 2000, (max intensity value is 4095, 12-bit CMOS sensor), and for $I_{12}$ ratio of 1.2, we calculate the uncertainty to be:

$$U_{I_{12}} = \sqrt{\left(\frac{1}{2000} 0.5\right)^2 + \left(-\frac{1.2}{2000} 0.5\right)^2}$$

$$U_{I_{12}} \approx 0.0003$$

For lower intensity values, say 200, the $U_{I12}$ was found to be 0.003, which is also a considerably low uncertainty value.

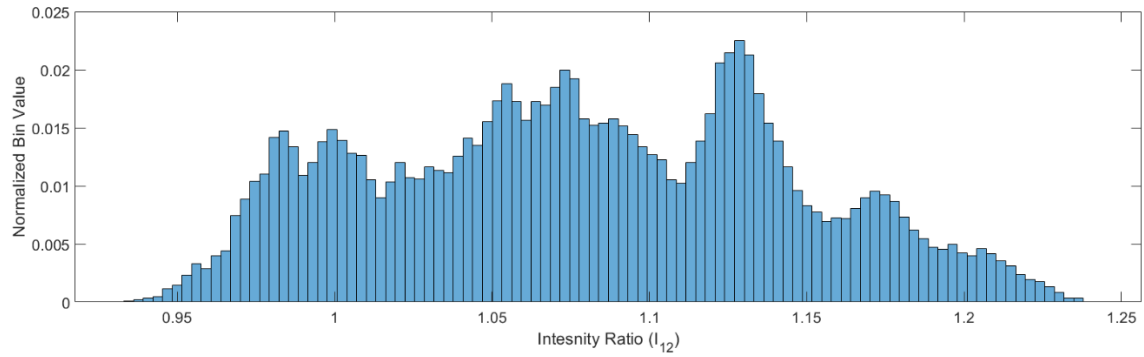

Figure S11 Normalized Histogram of the $I_{12}$ ratio for Layer 100 in the experiment of applying STWIP to monitor LPBF printing of fatigue bars. The bin values were normalized by the total count, i.e., Bin value = Bin Count/ Total Count.

**SI Section VI: Experimental Case of Applying STWIP to LPBF Process Measurement**

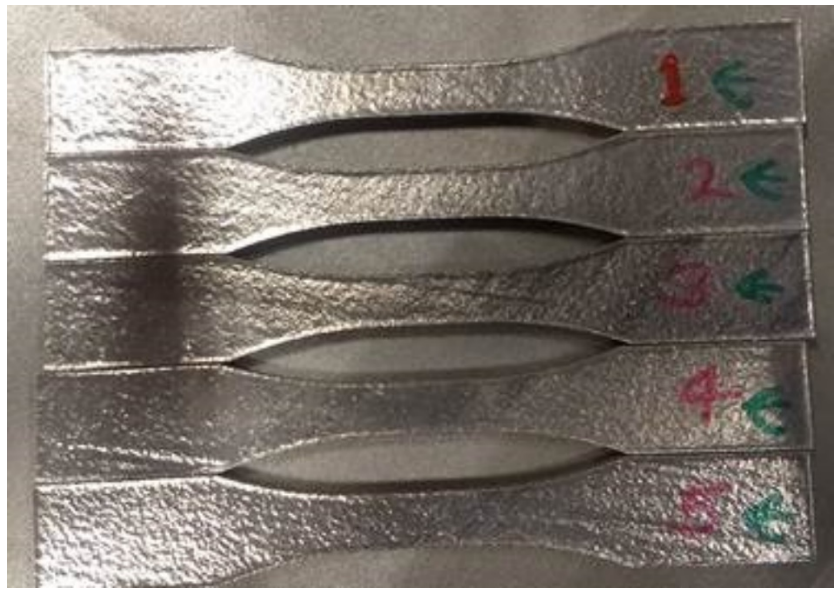

Figure S12: Photograph of the five printed Fatigue Test samples corresponding to the data and analysis presented in this work

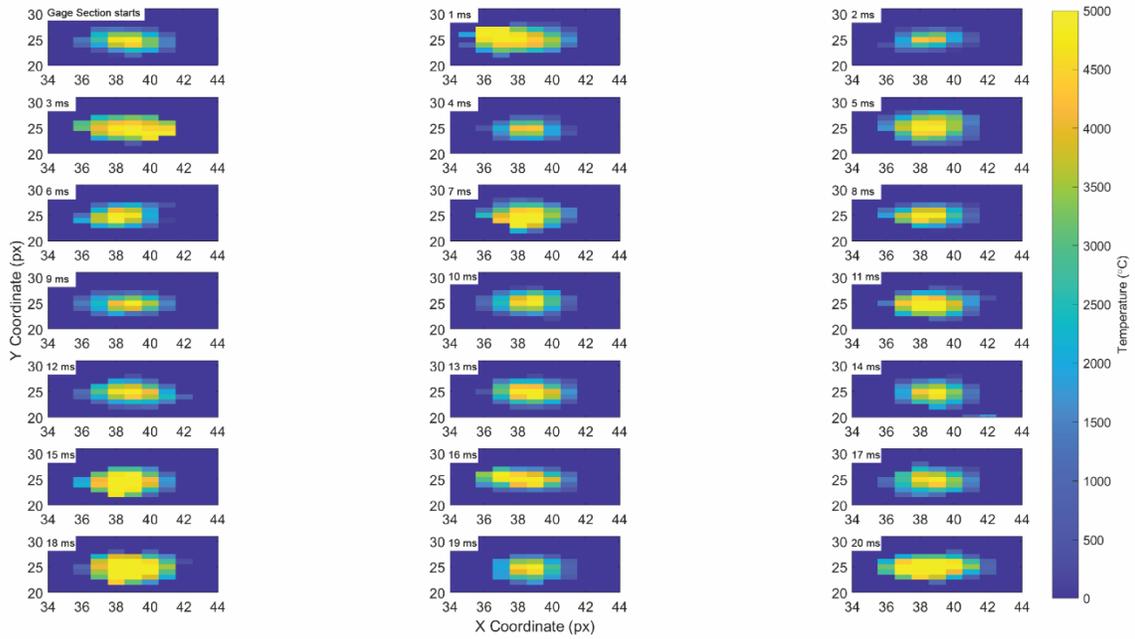

(a)

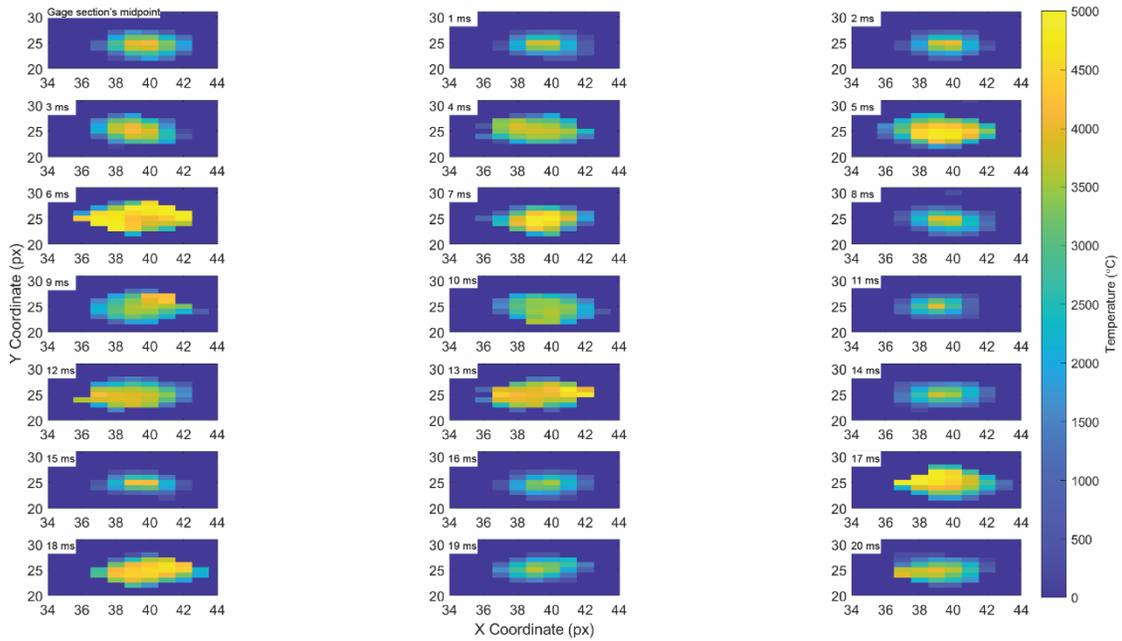

(b)

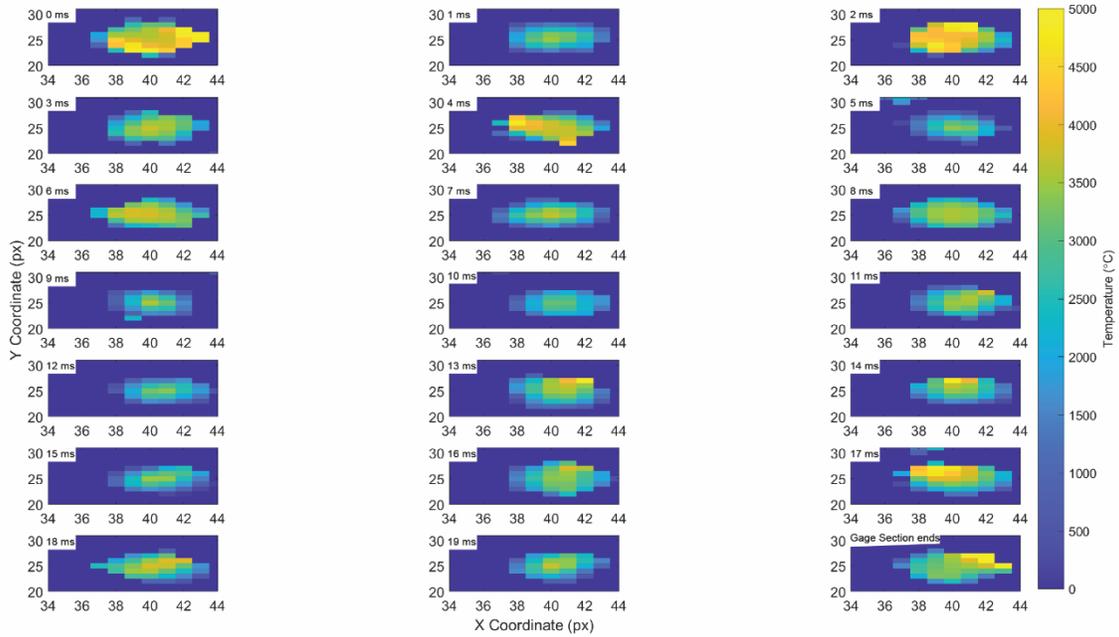

(c)

Figure S13 (a) the Beginning, (b) Middle and (c) End gauge region's MP core's temperature profile measured by the STWIP for Layer 60 in a test sample (Fatigue Bar 3), temporally separated by 1ms. Note: the blue pixels surrounding the detected MP boundary are pseudo background pixels and do not necessarily mean 0°C. [please refer to web version of the article for color images].

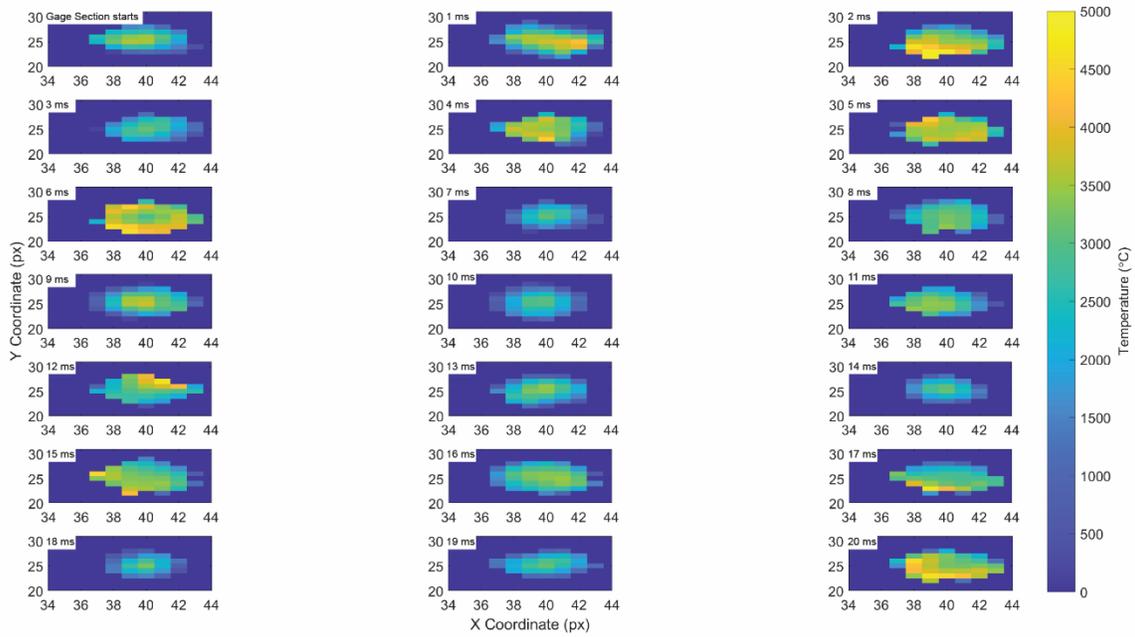

(a)

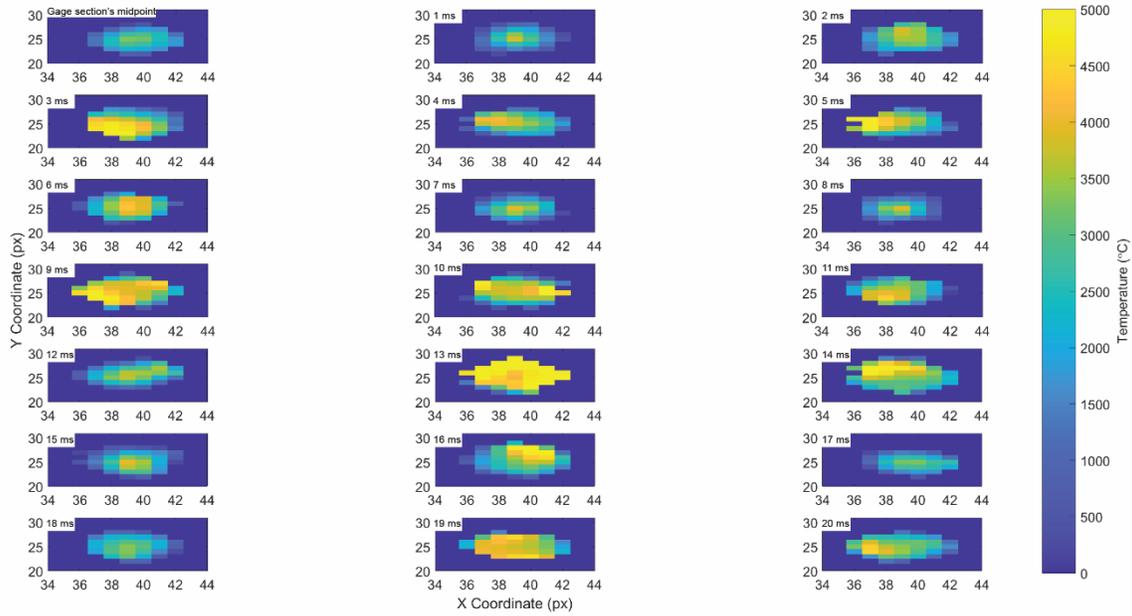

(b)

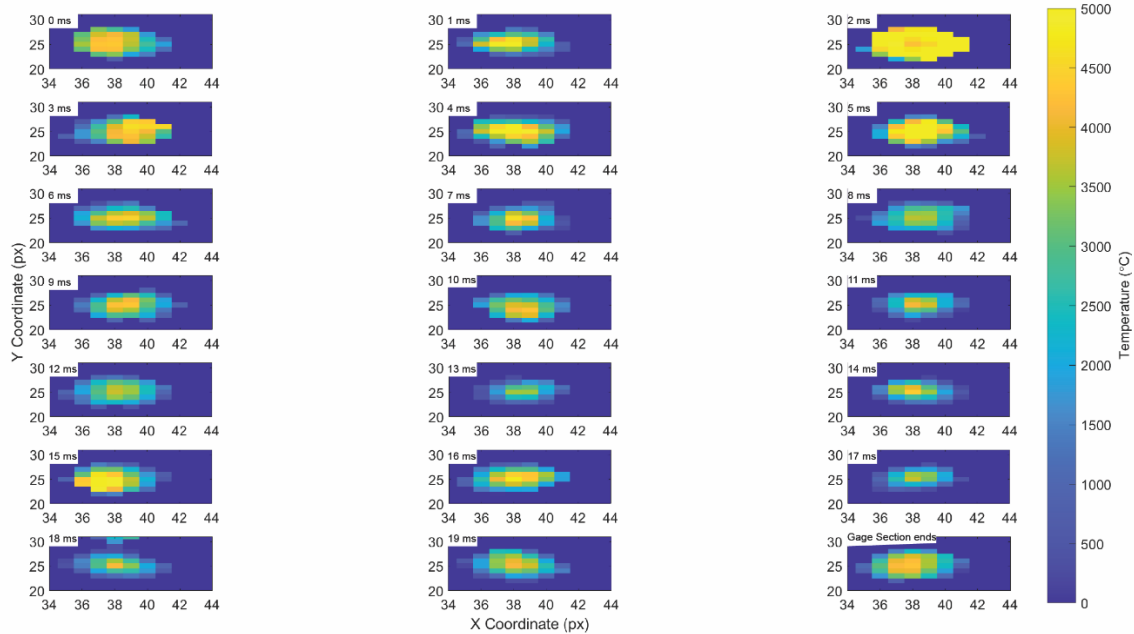

(c)

Figure S14: (a) The Beginning, (b) Middle and (c) End, gage section's MP core's temperature profile measured by the STWIP for Layer 100 in a test sample (Fatigue Bar 3), temporally separated by 1ms. Note: the blue pixels surrounding the detected MP boundary are pseudo background pixels and do not necessarily mean 0°C. [please refer to web version of the article for color images].

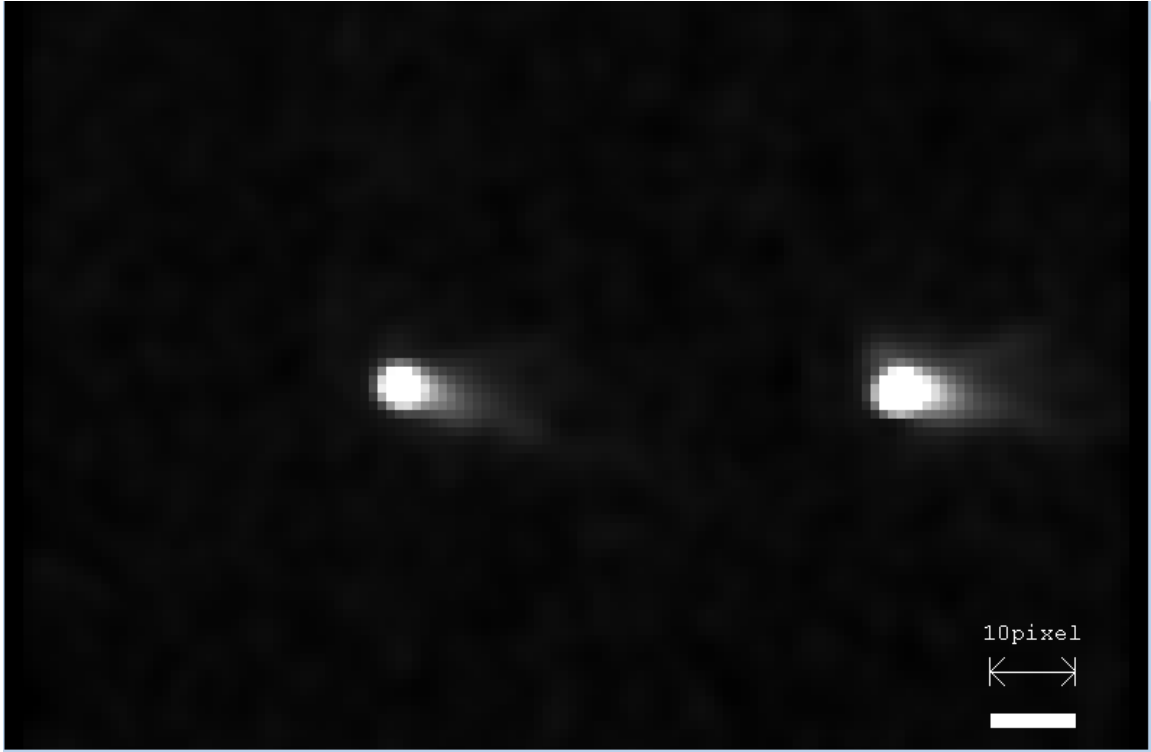

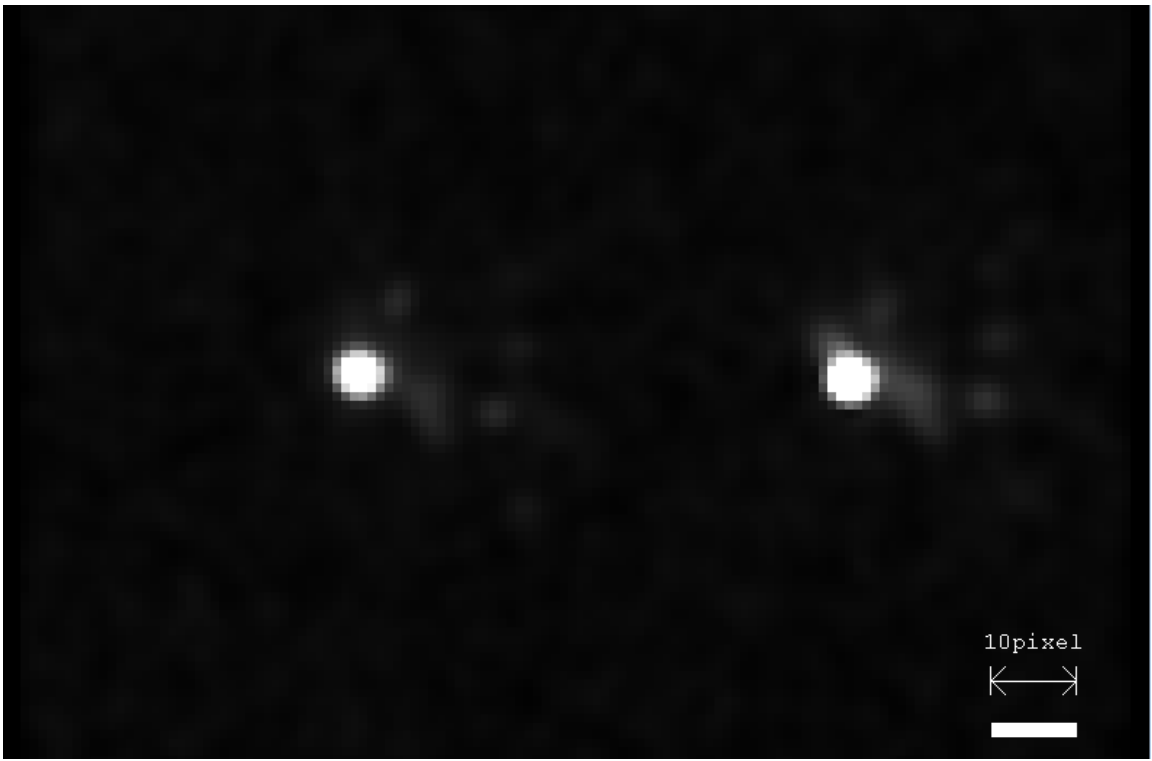

Figure S15: Representative melt pool images showing the melt pool length and spatter. These image correspond to a IN718 build at default print parameters (285W, 1000 mm/s). Pixel Scale – 1 pixel = 20μm